\documentclass{aa}  

\usepackage{graphicx}
\usepackage{txfonts}
\usepackage{natbib}
 \bibpunct{(}{)}{;}{a}{}{,} % to follow the A&A style
\def\apjl{ApJL }

\def\apjs{ApJS }

\def\aap{A\&A }

\def\cent{\mathrm{cm}}

\begin{document}

   \title{Dust absorption and scattering in the silicon K-edge}
   \titlerunning{The silicon K-edge}

   \subtitle{}

   \author{S.T. Zeegers
          \inst{1,2,6},
          \,
          E. Costantini\inst{1,4},
          \,
          D. Rogantini\inst{1},
          \, C.P. de Vries\inst{1},
          \,
          H. Mutschke\inst{3},
          \,
          P. Mohr\inst{3},
          \,
          F. de Groot\inst{5},
          \,
          A.G.G.M. Tielens\inst{2}          
          }
    \authorrunning{S.T. Zeegers, E. Costantini, D. Rogantini, C.P. de Vries}

   \institute{SRON Netherlands Institute for Space Research,
              Sorbonnelaan 2, 3584 CA Utrecht\\
              \email{S.T.Zeegers@sron.nl}
         \and
         Leiden Observatory, Leiden University, PO Box 9513, NL-2300 RA Leiden, the Netherlands
         \and
         Astrophysikalisches Institut und Universit\"{a}ts-Sternwarte (AIU), 
         Schillerg\"{a}\ss{}chen 2-3, 07745 Jena, Germany
         \and
         Anton Pannekoek Astronomical Institute, 
         University of Amsterdam, P.O. Box 94249, 1090 GE Amsterdam, The Netherlands
         \and 
         Debye Institute for Nanomaterials Science, 
         Utrecht University, Universiteitsweg 99, 3584 CG Utrecht, Netherlands
         \and
         Academia Sinica, Institute of Astronomy and Astrophysics, 11F Astronomy-Mathematics Building, NTU/AS campus, No. 1, Section 4, Roosevelt Rd., Taipei 10617, Taiwan
  }

   \date{Received 14 January 2019; Accepted 2 April 2019}

% \abstract{}{}{}{}{} 
% 5 {} token are mandatory
 
  \abstract
   {The composition and properties of interstellar silicate dust are not well understood. 
   In  X-rays, interstellar dust can be studied in detail
   by making use of the fine structure features in the Si K-edge.  
   The features in the Si K-edge offer a range of possibilities to study silicon-bearing dust, 
   such as investigating the crystallinity,  abundance, and the chemical composition along a given line of sight.}
   {We present newly acquired laboratory measurements of the silicon K-edge of several silicate-compounds 
   that complement our measurements from our earlier pilot study. The resulting dust extinction profiles serve as 
   templates for the interstellar extinction that we observe. 
   The extinction profiles were used to model the interstellar dust in the dense environments of the Galaxy.}
   {The laboratory measurements, taken at the Soleil synchrotron facility in Paris, 
   were adapted for astrophysical data analysis and implemented in the SPEX spectral fitting program. 
   The models were used to fit the spectra of nine low-mass X-ray binaries located in the Galactic center neighborhood
   in order to determine the dust properties along those lines of sight.}
   {Most lines of sight can be  fit well by amorphous olivine. 
   We also established upper limits on the amount of crystalline material that the modeling allows. 
   We obtained values of the total silicon abundance, silicon dust abundance, and depletion along each of the sightlines. We 
   find a possible gradient of $0.06\pm0.02$ dex/kpc for the total silicon abundance versus the Galactocentric distance.
   We do not find a relation between the depletion and the extinction along the line of sight.}
   {}

   \keywords{dust, extinction --
   X-rays: binaries -- ISM:abundances
                }

   \maketitle
%
%________________________________________________________________

\section{Introduction}\label{sec:introduction}
 
Silicates are an important and abundant component of  interstellar dust~\citep{Tielens01}.
They are found at every evolutionary stage in the life cycle of stars, such as interstellar clouds,  
the circumstellar environment of oxygen-rich asymptotic giant branch (AGB) stars, and protostellar disks, and are also found in  meteorites, in    comets, 
on Earth, and on other planets~\citep{Henning10}. 
It is therefore crucial to understand the composition and properties of the silicate dust
in order to make correct assumptions about each of the many processes in the universe where dust plays 
a role. 

Observations of the gas abundances in interstellar environments have given an indication of the dust composition in the Galaxy. 
From observations of the Sun, nearby stars, and the solar system it is known which elements are expected to be abundant in the interstellar medium (ISM).
 However, certain elements are depleted from the cold gas phase and are assumed  reside in dust particles~\citep{Draine03}. 
Here we define depletion as the ratio of the dust abundance to the total amount of a given
element (i.e., both gas and dust). A large fraction of the abundant elements carbon, oxygen, silicon, iron, and magnesium are thought to be depleted in dust
~\citep{Jenkins09,Savage96}. 
These elements form the basic constituents of silicates, except for carbon, which forms its own grain 
population of carbonaceous dust~\citep{Weingartner01}.  
However, the precise composition of interstellar silicates remains unknown.
The bulk of the interstellar silicate dust is thought to consist of olivine and pyroxene types of dust, with iron and magnesium,
and smaller amounts of  less abundant elements, such as calcium~\citep{Tielens01}.
In addition, there may be oxides present (e.g., SiO, SiO$_2$) since they are observed in stellar spectra~\citep{Posch02,Henning95},
and silicon could also be present in small amounts in the form of silicon carbide~\citep[SiC,][]{Kemper04,Min07}.

Much of our knowledge of interstellar silicates comes from the study of infrared absorption and emission features.
In the infrared, silicates are mainly observed by using the Si-O bending and stretching modes at 10 and $20\,\mu\mathrm{m}$.  
\citet{Draine84} used synthetic optical functions in combination with the observational opacities of the
$9.7\,\mu\mathrm{m}$ feature to derive an average composition of the silicate dust called astrosilicate, 
finding an olivine type with a composition of $\mathrm{Mg}_{1.1}\mathrm{Fe}_{0.9}\mathrm{SiO}_4$;  in this way the silicate grains would incorporate 
$90\%$, $95\%$, $94\%$, and $16\%$ of the total Si, Mg, Fe, and O, respectively. Nonetheless, 
interstellar silicates are expected to form a mixture of different silicate types, including olivine, but also pyroxene, for example.  
Studies by~\citet{Kemper04},~\citet{Chiar06}, and~\citet{Min07} compared laboratory spectra with observations and
find that a combination of olivine and pyroxene dust models fit the $10\,\mu\mathrm{m}$ feature, 
although they find varying amounts of olivine and pyroxene, and varying ratios of iron to magnesium.

An important property of the dust is the crystallinity, which can teach us about the formation history of the dust.
However, the formation process of crystalline dust is not well understood~\citep{Speck11}. 
We know from observations of circumstellar dust that evolved  oxygen-rich AGB stars produce silicate dust, and that 
up to about $15\%$ of this dust is in crystalline form. 
The dust is then subsequently injected in the ISM by stellar winds~\citep{Sylvester99,Kemper04}. 
Crystalline dust is thought to be formed close to the star and is expected to be mostly magnesium rich.  
Farther from the star the temperatures are lower and the silicates do not get the opportunity 
to crystallize~\citep{Molster102}. However, this formation process of crystalline dust is not certain. 
\citet{Speck15} found crystalline dust at the outer edge of the star HD 161796 
and amorphous dust in the inner part of the dust shell. They propose that crystallization may happen when the dust encounters the ISM.
Farther away from dust forming stars, in the diffuse ISM,
the smooth nature of the $\sim10\,\mu\mathrm{m}$ and $\sim20\,\mu\mathrm{m}$ features indicate that 
most of the interstellar dust is amorphous. 
Only $1.1\%$ of the dust along the lines of sight toward carbon-rich Wolf-Rayet stars near the center of the Galaxy 
appears to be crystalline~\citep{Kemper04}. 
From the mass budget of stellar dust sources  
the amount of crystalline dust in the ISM is expected to be $\sim5\%$~\citep{Kemper04}, which suggests considerable
re-processing of the dust in the ISM.
Over time the silicates may lose their crystalline structure in the violent environment of the ISM, where the dust is bombarded 
by radiation and cosmic rays on a timescale of $\sim70$ Myr~\citep{Bringa07}. These processes may be the cause of the amorphization
of crystalline dust in the ISM~\citep[][and references therein]{Jager03}.
Interestingly, silicates are again abundant in crystalline form in protoplanetary disks. The cores of some interstellar grains
retrieved by the stardust mission, also show the presence of crystalline dust~\citep{Westphal14}, 
showing the possibility that at least some of the crystalline interstellar dust may 
survive in the ISM. 

The soft X-ray part of the spectrum provides an alternative wavelength range
for the study of interstellar dust~\citep{Draine03,Lee09,Costantini12}.
X-ray binaries are used as a background source to observe the intervening gas and dust along the line of sight.
In the spectra of these X-ray binaries, we can observe several absorption edges. Depending on the column density along the line of sight and 
the brightness of the source, it is possible to access the absorption edges of different elements.  
The X-ray absorption fine structure (XAFS) near the atomic absorption edges of an element
can be used as a unique fingerprint of the dust along the line of sight toward the source.
XAFSs are observed in X-ray spectra of various astrophysical sources of \textit{Chandra} and XMM-\textit{Newton} 
\citep[][hereafter Z17]{Lee01,Ueda05,Kaastra09,Devries09,Pinto10,Pinto13,Costantini12,Valencic13,Zeegers17}.
They provide a powerful tool for the  study of the composition, abundance, crystallinity, stoichiometry, and size of interstellar
silicates~\citep[][Z17]{Devries09}.

Interstellar dust may show slight variations in different environments. For example, from dense molecular clouds we know 
that dust may incorporate a layer of ice around the grains, 
but  in less dense environments the dust may also show variations in composition and properties. 
For instance, the abundances of several elements show a decrease with distance from the Galactic center~\citep{Rolleston00,Chen03,Davies09}. 
The ISM is also known to be patchy and 
therefore may allow the observation of local differences in the chemistry along different lines of sight~\citep{Bohlin78,Nittler05}. 
The X-rays provide the possibility to study dust in different environments. 
In this study, we focus on the silicon K-edge, 
which gives access to denser regions in the central part of the Galaxy.

In our pilot study, we showed the analysis of the silicon K-edge of GX 5-1 (Z17). 
Here we expanded the number of sources  studied to a total of 9, and  we expanded the set of silicate samples with respect to the previous study from 6 to 15. 
The sample set contains the interstellar dust analogs pyroxenes, olivines, and silicon dioxide,
which can be used to analyze interstellar silicate dust. 
These measurements are part of a larger laboratory measurement campaign~\citep{Costantini13}. 

The paper is structured in the following way. 
In Section~\ref{abs_edges_ch3} we explain the analysis of our laboratory data and the use of XAFS to investigate the composition
of interstellar dust. In Section~\ref{ext_cross_ch3} we explain how we obtained the extinction cross section  and implemented 
them in the extinction profiles that can be used as interstellar dust models in an X-ray fitting code. 
In Section~\ref{data_an_lmxbs_ch3} we show the source selection, data, and spectral analysis.
In Section~\ref{discussion_ch3} we discuss the results and the chemistry of the dust toward the dense central area of the Galaxy. 
Lastly, in Section~\ref{summ_concl_ch3}, we give a summary and our conclusion. 

%__________________________________________________________________

\section{X-ray absorption edges}\label{abs_edges_ch3}

\subsection{Dust samples}\label{the_samples_ch3}
In this analysis, we make use of 14 different dust samples for which we measured the Si K-edges.
The composition and structure of these dust samples are listed in Table~\ref{table:samples}.
Samples 1-5  are the same as those used in Z17;  
samples, 6 - 14, were measured in January 2017 at the Soleil synchrotron facility in Paris.

There are several olivine- and pyroxene-type silicates among the samples, as well as different types of quartz. 
Although technically quartz types (samples 13-15) are oxides, we  refer to all the samples in the paper as silicates for simplicity.
Samples 2, 3, 5, 6, 7, 8, and 10 were synthesized for this analysis in laboratories at AIU Jena and Osaka University. 
In particular, the amorphous samples (2, 5, 7, 8, and 10)
were synthesized by quenching  a melt according to the procedure described in~\citet{Dorschner95}. 
There are also synthesized crystalline samples in the sample set, such as fayalite (sample 9). 
The crystals in this sample were grown via the 
`\textquoteleft scull method'\textquoteright~\citep{Lingenberg86}. 
More details about samples 1-5 can be found in Z17, and more details about samples 6-15 can 
be found in Table~\ref{table:samples}.

The samples were  chosen because of their relevance regarding the possible components of the silicate dust in the ISM.
We used the following criteria in the selection of this sample set: 
\begin{itemize}
\item The sample set consists  of pyroxenes, olivines, and oxides; 
\item The silicate samples have different iron-to-magnesium ratios; 
\item The samples contain both amorphous and crystalline silicates.
\end{itemize}

The motivation for these selection criteria is based primarily on \citet{Draine84}, 
who derived a general composition of  ``astronomical silicate''  based upon infrared emissivities inferred from observations of
circumstellar and interstellar grains. The composition of this silicate is used as a starting point for the composition of the silicates 
in our sample set. The content of the sample set is then further refined by involving the results of 
detailed studies of the 9.7 and $18\,\mu$m features in the infrared, which give an indication of the silicate dust composition. 
Since we expect amorphous silicates to be abundantly present~\citep{Kemper04}, 
the sample set contains seven amorphous samples, of which six have a different composition: one olivine, 
two quartz types, and four pyroxenes. 
From studies in the $8 - 13\,\mu$m band it can be concluded that a mix of predominantly pyroxenes and olivines fit 
the observed spectra well~\citep{Kemper04,Chiar06,Min07}. While~\citet{Kemper04} find that olivine dust dominates 
by mass in the ISM,
\citet{Chiar06} find that pyroxene dominates. Both pyroxenes and olivines are therefore well represented in our sample set.  
Furthermore, different values of the Mg-to-Fe ratio have been found in silicate dust in the ISM~\citep{Kemper04,Min07}. %refs
In order to be able to investigate this ratio, we used dust samples with different iron-to-magnesium ratios. In the case
of olivine-type silicates we explore both extremes, namely fayalite and forsterite.  
Fayalite is the iron endmember of the olivine group, whereas forsterite is the magnesium endmember.
Our sample also contains an amorphous olivine with equal amounts of iron and magnesium.   

\begin{table}
\caption{Samples} % title of Table
\label{table:samples} % is used to refer this table in the text
\centering % used for centering table
\begin{tabular}{l l l l} % centered columns (4 columns)
\hline\hline % inserts double horizontal lines
No. & Name &Chemical formula& Structure  \\ % table heading
\hline \noalign{\smallskip}% inserts single horizontal line
1 & Olivine&$\mathrm{Mg}_{1.56}\mathrm{Fe}_{0.4}\mathrm{Si}_{0.91}\mathrm{O}_4$& crystal  \\ % inserting body of the table
2 & Pyroxene&$\mathrm{Mg}_{0.9}\mathrm{Fe}_{0.1}\mathrm{Si}\mathrm{O}_3$& amorphous  \\
3 & Pyroxene&$\mathrm{Mg}_{0.9}\mathrm{Fe}_{0.1}\mathrm{Si}\mathrm{O}_3$& crystal \\
4 & Enstatite&$\mathrm{Mg}\mathrm{Si}\mathrm{O}_3$& crystal$^*$   \\
5 & Pyroxene&$\mathrm{Mg}_{0.6}\mathrm{Fe}_{0.4}\mathrm{Si}\mathrm{O}_3$& amorphous\\
6 & Pyroxene &$\mathrm{Mg}_{0.6}\mathrm{Fe}_{0.4}\mathrm{Si}\mathrm{O}_3$& crystal\\
7 & Olivine &($\mathrm{Mg}_{0.5}\mathrm{Fe}_{0.5})_2\mathrm{Si}\mathrm{O}_4$& amorphous\\
8 & Pyroxene &$\mathrm{Mg}_{0.75}\mathrm{Fe}_{0.25}\mathrm{Si}\mathrm{O}_3$& amorphous\\
9 & Fayalite &$\mathrm{Fe}_{2}\mathrm{Si}\mathrm{O}_4$& crystal\\
10 & Enstatite &$\mathrm{Mg}\mathrm{Si}\mathrm{O}_3$& amorphous\\
11 & Forsterite &$\mathrm{Mg}_{2}\mathrm{Si}\mathrm{O}_4$& crystal\\
12 & Quartz &$\mathrm{Si}\mathrm{O}_2$& crystal\\
13 & Quartz &$\mathrm{Si}\mathrm{O}_2$& amorphous\\
14 & Quartz &$\mathrm{Si}\mathrm{O}_2$& amorphous\\
\hline %inserts single line
\end{tabular}
\tablefoot{$^*$Sample 4 contains a very small amount of iron, which is not significant in our analysis. The Fe-to-Mg ratio is $4\times10^{-2}$. 
More information about samples 1-5 can be found in Z17.
Sample 6 is a pyroxene, and is the crystalline counterpart of sample 5. 
Sample 7 is an amorphous olivine with equal contributions of iron and magnesium.
Sample 8 is an amorphous pyroxene with an Fe-to-Mg ratio of 1:3.
Sample 9, fayalite, was synthesized at the   University of Frankfurt, Physical
Institute~\citep{Fabian01}. 
Sample 10 is an amorphous enstatite  synthesized at AIU Jena. 
Sample 11, forsterite, is a commercial product of Alfa Aesar.
Sample 12 is a natural rock crystal from Brazil~\citep{Zeidler13}.
Sample 13, an amorphous silica, is a commercial product of  Qsil Ilmenau, Germany, named 
\textquotedblleft ilmasil\textquotedblright.
Sample 14 is a commercial amorphous silica powder supplied by   Fisher Scientific.
Samples 13 and 14 differ in degree of amorphization (see the XAFS,  lower right panel of
Figure~\ref{fig:edges_again}).}
\end{table}

\subsection{Analysis of laboratory data}\label{ana_lab_ch3}

A self-evident method to measure the degree of absorption of a sample would be to measure the transmission of the radiation 
through the sample. This can be done by measuring the ratio of the intensity of the incoming beam to that of the transmitted beam. 
In order to appear optically thin at the energy around the Si K-edge, this 
measurement would require a sample thickness of $1.0 - 0.5 \,\mu\mathrm{m}$. 
It is impractical to perform measurements with such thin samples at X-ray energies. 
Therefore, the degree of absorption of the samples cannot be measured directly through transmission. Instead,
the absorption is derived from the fluorescent measurements of the Si
K$\alpha$ line in our analysis of the Si K-edge. 
An overview of the theory behind this method can be found in Section 3.2 of Z17. 

The fluorescent measurements of the samples were performed at the SOLEIL synchrotron facility 
in Paris using the~Line for Ultimate Characterisation by Imaging and Absorption (LUCIA) ~\citep{Flank06}.
The first run was completed and published in Z17, the second run was completed in 2017 and presented in this paper.
All the samples were pulverized into a fine powder. This powder was then pressed into a layer of indium, 
which was adhered to a copper sample plate.
The sample plate was placed in the X-ray beam in a vacuum environment. The reflecting fluorescent
signal was measured by a silicon drift diode detector. Each sample was measured at two different positions 
on the sample plate and for each position the measurement was repeated once, resulting in a total of four measurements per sample. 
The change in position was necessary to avoid any dependence of the measurement on the position of the sample on the copper plate.
We took the average of the resulting four measurement.
In order to obtain the noise in the measurements, we determined the dispersion among the measurements.
We found a small dispersion of $5\%$. 
This is slightly higher than in our previous sample set of Z17,
but still considered negligible. 
Before the start of the measurement run, the instrument was calibrated for the Si K-edge energy. All the measurements on this
edge were performed on the same day to ensure that each of the measurements was performed under the same conditions.
The energy calibration of the instrument is important to avoid discrepancies between the observations and the measurements~\citep{Gorczyca13}.
We compared the measurement of the crystalline quartz (sample 12), to three other measurements from the literature in order to check our 
calibration on the absolute energy scale~\citep{Li1995,Nakanishi_2009,Ohta17}; we found  a perfect match among the measurements.
We are therefore confident that the uncertainty on the energy scale is modest and smaller than the energy resolution of \textit{Chandra}.
After the measurements are obtained they need to be corrected for two effects, namely pile-up and  saturation. 
Pile-up occurs when two photons hit the detector at once and are recorded as one event with double the energy. This 
effect can be seen in the spectrum as a spurious line appearing at twice the expected energy. This effect is, however, minimal ($<1\%$).
Saturation occurs because the measured fluorescent signal is not directly proportional
to the absorption. The effect of saturation becomes important when a sample is concentrated, which is the case here~\citep{Groot12}.  
We use the program FLUO to correct for saturation when needed~\citep{Stern1995117}.
More details can be found in Z17;  here we follow the same correction procedure.
We note that a crystalline pyroxene was omitted from our sample set; it is present in Z17 and is indicated there as sample 6. 
The correction is on average three times larger than in the other samples, which indicates that the sample could be overcorrected by the application of FLUO.
The overcorrection may be the result of a mismatch between the given composition and the
measured compound, although this may not necessarily be the case~\citep{Ravel05}. 
However, since the cause of the large correction could not be clarified in this case, the sample was removed from the sample set.

\subsection{X-ray absorption fine structures}\label{fine_struct_ch3}

X-ray absorption fine structures are modulations that arise when an X-ray photon 
excites a core electron in an atom. A modulation is a fingerprint of one type of dust and can therefore be used
to discriminate between different types of dust in the ISM.  
XAFSs arise from the wave-like nature of the photoelectric state. 
When a core electron is excited by an incoming X-ray with the right energy,  
the ionized electron will then behave like a photoelectron. 
This can be interpreted as a wave emanating from the site of the absorbing atom.
Depending on the available energy, the photoelectron 
can scatter around the neighboring atoms. Due to this interaction, the initial wave is scattered and new waves emanate from 
the neighboring atoms. These waves are superimposed on the original wave creating interference. 
This subsequently changes the probability of the photoelectric effect. We observe the constructive 
and deconstructive interference in the edge as a function of energy, i.e., the XAFS. 
Depending on the elements and the position of the neighboring 
atoms, the XAFS are modified in a unique way, reflecting the crystallinity and chemical composition of the material. 
\begin{figure*}
 \begin{center}
  \advance\leftskip-1cm
 \includegraphics[scale=0.7]{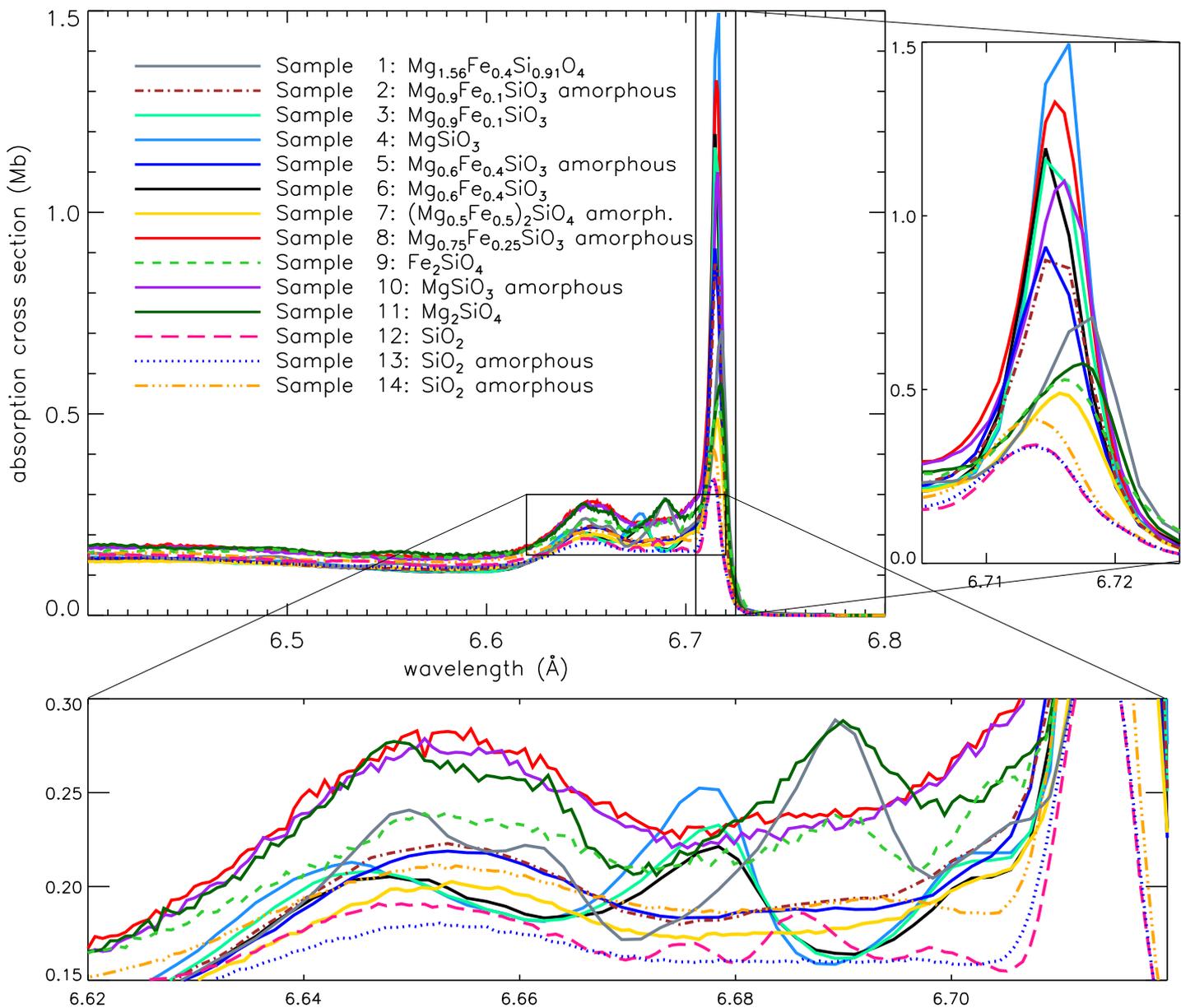}
 \caption{\small{The Si K-edge of dust samples 1-14 (samples 1-5 are from Z17). The x-axis shows the energy in $\text{\AA}$ and the 
 y-axis shows the amount of absorption indicated by the cross section (in Mb per Si atom).}}
 \label{fig:all_edges}
  \end{center} 
\end{figure*}

The resulting absorption cross sections can be found in Figure~\ref{fig:all_edges}. 
We highlight the XAFS in the  inset on the left. Samples 2, 5, 7, 8, 10, 13, and 14 are amorphous. 
Since these amorphous samples do not have a regular crystalline
structure, they lose the distinct signature that is present in their crystalline counterparts. This effect can be
observed in Figure~\ref{fig:all_edges} 
between $6.67$ and $6.70\,\text{\AA}$.
The XAFS of amorphous materials are very similar in shape (compare  samples 7, 8, and 10). 
It is therefore difficult to distinguish between  an amorphous pyroxene and an amorphous olivine. The differences 
between these edges will then also depend on slight shifts in the ionization energy and the peak intensity at this 
energy. This is shown by the inset  on the right in Figure~\ref{fig:all_edges}.
In the case of samples 12, 13, and 14 we can compare three samples where the compositions are the same, but the 
degree of crystallinity varies. It can be observed that the resulting absorption 
cross sections of quartz are very similar since the composition does not change from sample to sample, but that the crystalline
sample shows features between $6.67$ and $6.70\,\text{\AA}$. 
Distinct differences can be observed in the case of
the crystalline pyroxene of sample 7 and forsterite (sample 12) around $6.66$ and $6.70\,\text{\AA}$, 
illustrating the difference between crystalline pyroxene-type silicates and crystalline olivine-type silicates. 
The effect of the varying iron content of the samples is subtler and shows itself by 
shifts in the peak of the XAFS (e.g., samples 9 and 11). 
Furthermore, the samples can also be characterized by the peak strength of the edge between $6.71$ and $6.72\,\text{\AA}$ 
and the energy position of this peak.  

\section{Extinction cross sections}\label{ext_cross_ch3}

The extinction cross sections for each of the samples can be derived from the laboratory data. 
In this section, we will give a summary of the methods that are used to derive the cross sections. 
For a full description of the calculation of the extinction cross section,  we refer to Z17. 
From the laboratory data, we obtain the attenuation coefficient ($\alpha$). The Beer-Lambert law can be used
to derive $\alpha$:
\begin{equation}
 T=\frac{I}{I_0}=e^{-\alpha x} = e^{-x/l}.
 \label{eq:beer_lambert_ch3}
\end{equation}
Here $T$ is the transmittance, which can be obtained by assuming an optically thin sample thickness $x$ and by using tabulated
values of the mean free path $l$ (e.g., the average distance
travelled by a photon before it is absorbed) provided by the Center for X-ray Optics (CXRO) at
Lawrence Berkeley National laboratory. The laboratory absorption edges were transformed into transmission spectra and
fitted to the transmittance $T$ obtained from tabulated transmission data and from those provided by CXRO. 
Consequently, from $\alpha$ the imaginary part of the refractive index $k$ can be derived, since the attenuation coefficient
can be described as
\begin{equation}
\alpha=\frac{4\pi k}{\lambda}, 
\end{equation}
where $\lambda$ is the wavelength. 
The real part of the refractive index is then calculated by using a numerical solution to the Kramers-Kronig relations~\citep{Bohren10}. 
The method used for this calculation is the same as in Z17, namely the fast Fourier transform  (FFT) routines,
as described in \citet{Bruzzoni2002}. 
We used  Mie theory~\citep{Mie1908} to calculate the extinction efficiency at each wavelength and grain size. 
The grain size distribution used in this analysis is the MRN distribution, with a grain size range of $0.005 - 0.25\,\mu\mathrm{m}$.
The grains are modeled as solid spheres.
The MIEV0 code \citep{Wiscombe80} was used to calculate the extinction efficiency, which needs the optical constants, wavelength, and grain size as input parameters. 
From the obtained extinction efficiency we calculate the extinction cross sections. 
These extinction cross section are implemented in the Amol model of the fitting code SPEX~\citep{Kaastra1996}, 
where they are used for further analysis. Figure~\ref{fig:models} 
shows the resulting extinction profiles of each of the dust models. 
The absolute cross section are available in tabular form\footnote{\tiny{\url{www.sron.nl/~elisa/VIDI/}}}.

\section{Data analysis of the LMXB}\label{data_an_lmxbs_ch3}

\subsection{Source selection}\label{source_selec_ch3}
We selected nine low-mass X-ray binary sources for our analysis from the \textit{Chandra} Transmission Gratings Catalog
and Archive\footnote{\tiny{\url{http://tgcat.mit.edu/}}}. 
The selection depends on the brightness of the source and 
the hydrogen column density $N_\mathrm{H}$ towards the source. In order to have the best view of the silicon K-edge,
the column density of the source 
should be between $10^{22}$ and $10^{23}\,\cent^{-2}$. 
In addition, it is  important that the source be bright in order to observe the edge with a high signal-to-noise ratio.
The flux level needs to be
$>0.5\times10^{-12}\,\mathrm{erg}\,\mathrm{cm}^{-2}\mathrm{s}^{-1}$ at energies between 0.5-2 keV. 
The source should preferably not strongly fluctuate in brightness since this will
affect the quality of the edge in the spectrum. We therefore inspected the light curves for strong dips in the brightness. We did not
find this to be a problem in any of the selected sources. 
Sources with the desired column density lie preferentially around the Galactic center (GC) area 
(Table~\ref{table:sources}).

Another more practical selection criterion is that the source has to be observed in TE mode. 
The ACIS detectors on board Chandra can operate in different observing modes, 
namely continuous clocking (CC) mode and timed exposure (TE) mode. The CC mode is not suitable for measurements of the Si K-edge
since the edge is filled by the bright scattering halo radiation of the source. 
The edge has a different optical depth in comparison with the TE mode and seems slightly smeared.
The effect of the scattering halo is particularly evident in the CC mode because the 
two arms of the mode are now compressed into one. 
An overview of the sources that are used in this study is given in Table~\ref{table:sources}. Here we also indicate the observation IDs (obsids)
of the spectra of the sources, the distances, and the Galactic coordinates. 

\begin{table*}
\caption{Sources} % title of Table
\label{table:sources} % is used to refer this table in the text
\centering % used for centering table
\begin{tabular}{l c c c c} % centered columns (4 columns)
\hline\hline % inserts double horizontal lines
Name & obsid(s) & distance & \multicolumn{2}{c}{coordinates} \\ % table heading
     &  & kpc  &  $l$ (deg) & $b$ (deg) \\
\hline % inserts single horizontal line
%\rule{0pt}{2.3ex}
GX 5-1 & 19449, 20119 &  9.2$^1$ & 5.08 & -1.02 \\ % inserting body of the table
GX 13+1 & 11814, 11815, 11816, 11817&  $7\pm1^2$ & 13.52 & +0.11\\
GX 340+00 & 1921, 18085,19450,20099 & $11\pm0.3^3$ & 339.59 & -0.08\\
GX 17+2  & 11088 & 12.6$^4$ & 16.43 & +1.28 \\
4U 1705-44 & 5500,18086, 19451, 20082 & $7.6\pm0.3^5$ & 343.32 & -2.34 \\
4U 1630-47 & 13714, 13715, 13716, 13717 & 10$^6$ & 336.91 & +0.25 \\
4U 1728-34 & 2748 & $5.2\pm0.5^7$ & 354.30 & -0.15 \\
4U 1702-429 & 11045 & 7$^8$ & 343.89 & -1.32 \\
GRS 1758-258 & 2429, 2750 &  8.5$^9$ & 4.51 & -1.36 \\
\hline %inserts single line
\end{tabular}
\tablefoot{Table with \textit{Chandra} observations used in this paper, indicated by the obsids.
All observations were done in TE mode. Coordinates were taken from the SIMBAD database~\citep{Wenger00}. Distances are taken from
\citet{Christian97}$^1$, \citet{Bandyopadhyay99}$^2$, \citet{Christian97}$^3$, \citet{Lin12}$^4$, \citet{Galloway08}$^5$, 
\citet{Parmar86,Augusteijn01}$^6$, \citet{Galloway08}$^7$, \citet{Oosterbroek91}$^8$, and \citet{Keck01}$^9$}
\end{table*}

\subsection{Modeling procedure}\label{model_proc_ch3}
After the selection of the sources as described in the previous section, we inspected the spectra for pile-up. 
Pile-up occurs when two or more photons are detected as one single event, and therefore often occurs  in the spectra of
bright sources such as the X-ray binaries used in this work. 
Both gratings (HEG and MEG) are affected, but the effect is especially evident in the MEG grating.
The parts of the spectra of the MEG grating that were too affected by pile-up were ignored. This was done in the case 
of all the X-ray binaries, but the ignored range varies per source. This is described for each individual source in
Appendix~\ref{appendix:data_tables_ch3}. 
Before we can study the interstellar dust, we first have to model the underlying continuum of each source and 
inspect the spectra for the presence of outflowing ionized gas and hot gas present along the line of sight,
as described in Sections~\ref{cont_neut_abs_ch3} and~\ref{hot_gas_ch3}.

\subsubsection{Continuum and neutral absorption}\label{cont_neut_abs_ch3}

The underlying continuum of each source is fitted using the spectral analysis code SPEX. 
We tested several additive SPEX models in our analysis, namely blackbody, disk-blackbody,
power-law, and Comptonization models.
The X-ray radiation from the source is then absorbed by a neutral absorber.
This model is represented by the multiplicative HOT model in SPEX. In order to mimic a neutral cold gas, 
the temperature of this gas is frozen to a value of kT = $5 \times 10^{-4}$ keV. 
After obtaining a  best fit for the continuum with absorption of neutral cold gas, we are able to determine 
the column density of hydrogen toward the source.

\begin{figure}
 \begin{center}
 \includegraphics[scale=0.5]{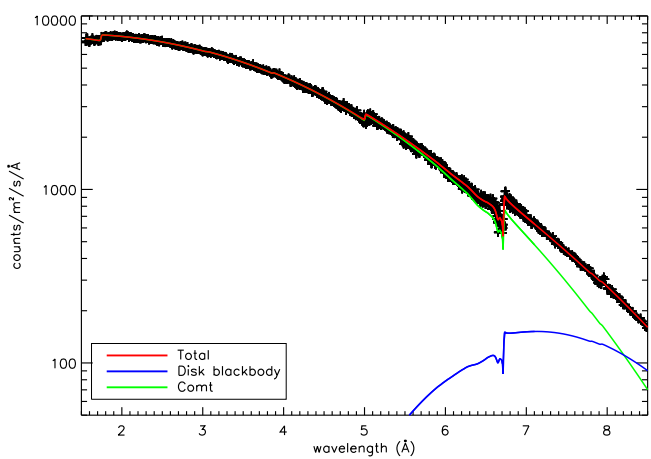}
 \caption{\small{Broadband fit of GX 5-1 with the data of obsid 19449 with the disk blackbody model (blue line) and 
 the Comptonization model (green line). Also shown is the total fit (red line). The AMOL dust models are included in the fit.}}
 \label{fig:broadband}
  \end{center} 
\end{figure}

We  use GX 5-1 as an example to explain our method of fitting the sources.
The results of the best fits of the other X-rays binaries and the details of the fitting method used in each case 
can be found in Appendix~\ref{appendix:data_tables_ch3}.
Two spectra were used in the fit of GX~5-1,  obsids 19449
and 20119. These observations have an excellent signal-to-noise
ratio; in the case of obsid 19449 this is around the Si K-edge
$S/N\approx100$ per bin, and for obsid 20119 $S/N\approx60$ per bin. The MEG grating shows signs of
pile-up in both data sets above an energy of 2.5 keV, and the data was therefore ignored above this value of the energy. 
Since the Silicon K-edge starts around 1.84 keV the MEG data is included in the analysis of the edge. This is the case for every X-ray binary in our analysis.
We used a Comptonization model and a disk-blackbody
model to describe the underlying continuum of GX~5-1. We found a column density of $5.8\pm0.2\times10^{22}\mathrm{cm}^{-2}$.
The broadband fit of GX~5-1 is shown in Figure~\ref{fig:broadband}. 
For clarity, the data displayed 
in this figure belongs to obsid 19449, since this data set dominates the fitting of the spectrum due to its superior quality.
The fit already includes the dust model (see Section~\ref{spex_ch3} for details). 
The parameter values of the best fit of GX~5-1 can be found in
Table~\ref{table:gx51}. Errors given on parameters are 1$\sigma$ errors, which is the case for all errors shown in this analysis.

\begin{figure*}
 \begin{center}
 \includegraphics[scale=0.6]{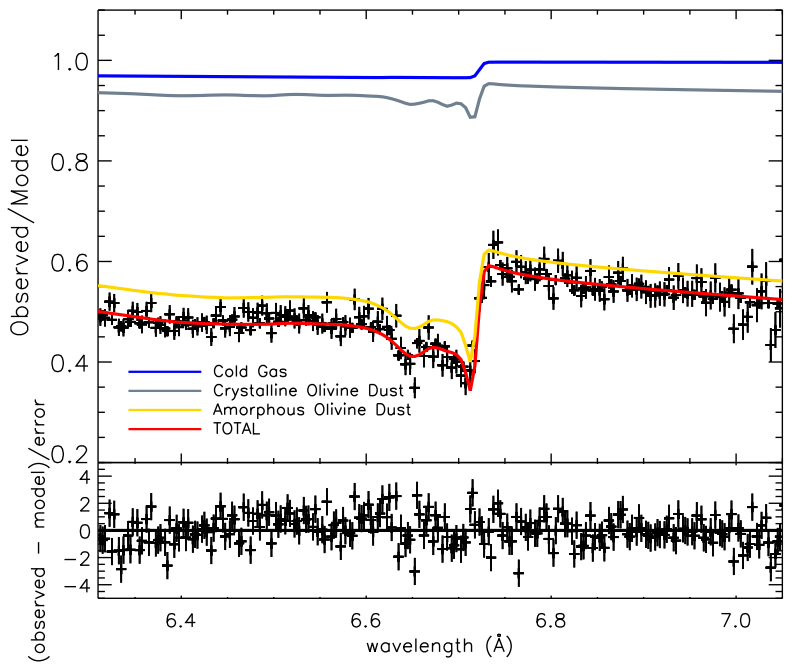}
 \caption{\small{Fit of the Si K-edge of X-ray binary GX 5-1. The best fitting dust mixture is shown by the yellow line (amorphous olivine, sample 8)
 and the gray line (crystalline olivine dust, sample 1). The cold gas contribution is shown by the blue line
 and the total (cold gas and dust) by the red line.}}
 \label{fig:best_fitgx5}
  \end{center} 
\end{figure*}

\subsubsection{Hot ionized gas on the line of sight in the Si K-edge region?}\label{hot_gas_ch3}

We tested whether there was hot gas along the line of sight towards the sources 
(fitted in the spectra using again the HOT model, which has a tuneable temperature), 
as well as outflowing ionized gas related to the source (fitted in the spectra using the XABS model of SPEX). 
If absorption lines of this gas appear near the edge, 
it is important to take these lines into account for accurate modeling.

We found evidence of gas intrinsic to the source in two cases, namely for GX 13+1 and 4U 1630-47. These sources 
show strong absorption lines in their spectra. In the case of GX 13+1, a second but non-outflowing  
ionized gas component was also found (Table~\ref{table:GX131}).  

Collisionally ionized hot absorbing gas in the ISM is thought to have temperatures between $\sim10^{6}-10^{7}\,\mathrm{K}$
~\citep{Yao07,Wang09,Wang13}.
For GRS 1758-258, GX 17+2, and 4U 1705-44 a hot component was found with temperatures within the range mentioned,
namely $1.5^{+0.6}_{-0.3}\times10^{7}\,\mathrm{K}$ , $1.6\pm0.2\times10^{6}\,\mathrm{K}$, and
$2.1^{+0.9}_{-0.6}\times10^{6}\,\mathrm{K}$, respectively. 
The hydrogen column density of the hot gas  for all sources is on the order of $10^{20}\,\mathrm{cm}^{-2}$
(Table~\ref{table:GRS1758258_GX172_4U170544}), which is in agreement with 
the typical expected hydrogen column densities~\citep{Yao05,Yao06, Yao07}.
We did not find evidence of outflowing ionized gas along the line of sight of GX 5-1, nor  did we find any contribution of hot gas.
This is also the case for 4U 1702-429, 4U 1728-34, and GX 340+00.
The values of the parameters of the HOT and the XABS model can be found in the tables with the best fits of the sources 
in Appendix~\ref{appendix:data_tables_ch3}.

\subsubsection{Dust mixtures and the SPEX AMOL model}\label{spex_ch3}

\begin{figure*}
 \begin{center}
 \includegraphics[scale=0.8]{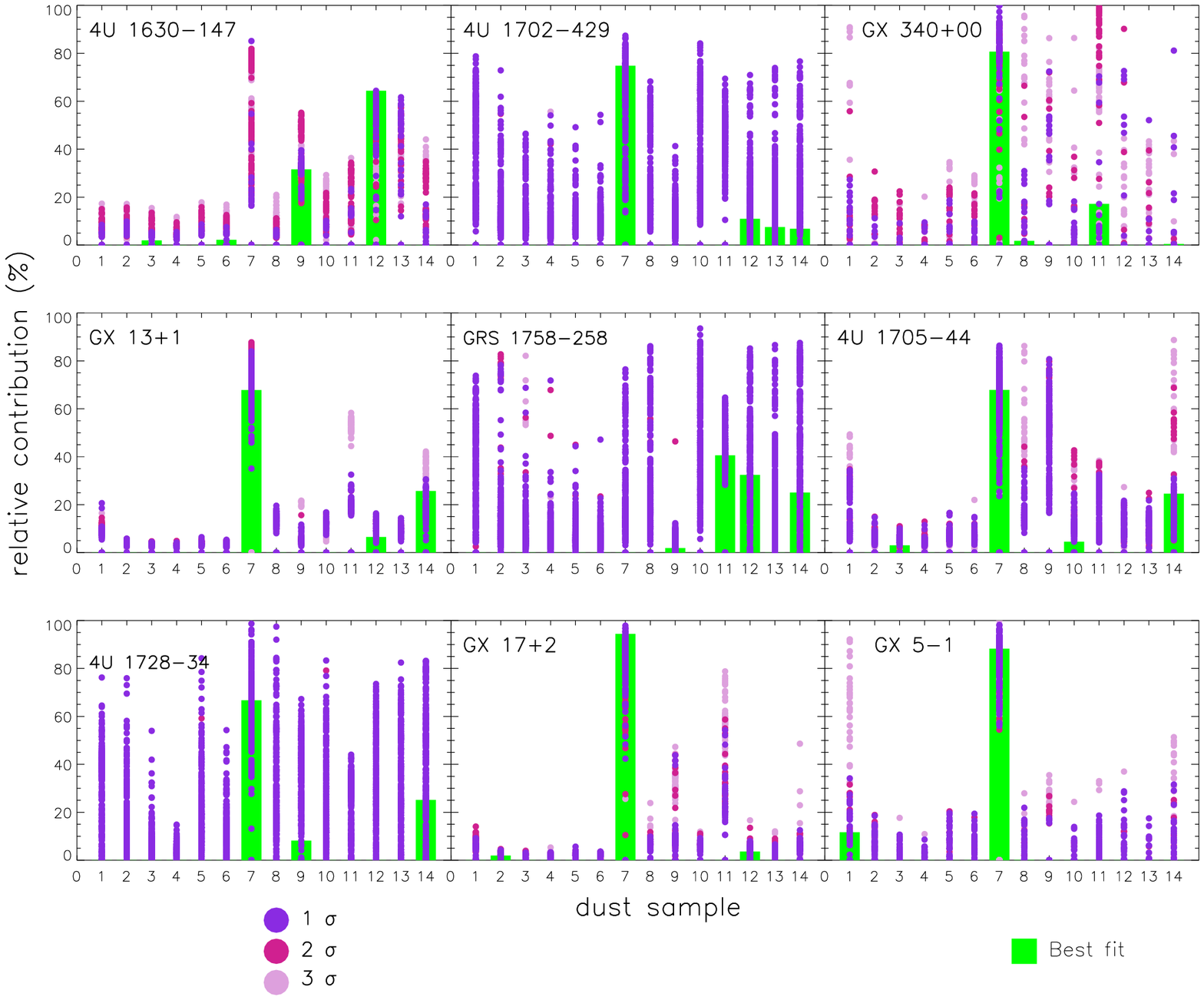}
 \caption{\small{Best fitting dust mixtures in comparison with other possible dust mixtures. 
 The relative contribution 
 of each of the dust components in the best fit is shown in green. The other dust mixtures are represented by the colored dots. 
 The colors indicate the 1, 2, and 3 $\sigma$ deviations from the best fitting dust mixture. For each dust mixture, four dust samples are fitted to the 
 spectra of the X-ray binaries.}}
 \label{fig:barchart}
  \end{center} 
\end{figure*}

After obtaining the underlying continuum and the column density of neutral hydrogen towards the source, 
we proceeded by adding the dust model to the fit. 
We also took into account the presence of hot and outflowing ionized gas along the line of sight, as mentioned 
in Section~\ref{si_abundance_depletion_ch3}. 
The AMOL routine in SPEX is used to fit the dust models to the data. AMOL can fit four dust models simultaneously. We wanted to test
all possible unique combinations of the 14 dust models. To execute these fits, we followed the method of~\citet{Costantini12},
namely the total number of fits ($n$) is given by $n=n_{\mathrm{dust}}!/(4!(n_{\mathrm{dust}}-4)!),$  where $n_{\mathrm{dust}}$ is the number 
of available dust models and 4 is the number of models that can be tested in the same run. 
This results in 1001 possible unique combinations to fit the 14 dust models.

From all these combinations we can select the best fitting mixture. 
Of each possible dust mixture, we determined
the reduced $C^2$ values. All the fits in this paper generated by SPEX are using C-statistics
\citep{Cash79} as an alternative to $\chi^2$-statistics. C-statistics can  be
used regardless of the number of counts per bin; therefore,
we can use bins with a low count rate in the spectral fitting. 
The best fit is given by the lowest reduced $C^2$ value~\citep{Kaastra17}. 
As an example, we show the resulting best fit of GX 5-1 in Figure~\ref{fig:best_fitgx5}. 
Here we show the contribution of cold gas,
and the two best fitting dust samples in the mixture: sample 1 crystalline olivine contributing $11\%$ and sample 7 amorphous olivine
contributing $89\%$ to the total column density of silicon in dust. The other two samples
in the mix do not contribute significantly. 
The best fits of the sources in our sample are described in Section~\ref{discussion_ch3}.

In Figure~\ref{fig:barchart} 
the best fitting dust mixtures of all nine X-ray binaries considered here are indicated by 
the green bars. On the x-axis we show the numbers belonging to each dust sample (see Table~\ref{table:samples}).
The y-axis indicates the relative contribution of each of the dust sample in the fit with respect to the dust column density of silicon. 
The best fits represent one of the 1001 possible dust mixtures per source, and it is useful to take the performance of the other 
dust mixtures into account before discussing the results. 
Therefore, an insightful way to study whether a certain dust mixture fits the edge well 
is by showing how much a dust mixture deviates from the best fitting mixture. 
Each of the 1001 possible dust mixture is represented in Figure~\ref{fig:barchart} 
by a set of four circles of the same color, 
i.e., the number of dust samples per fit. 
The position of the filled circles on the y-axis shows the contribution of the dust sample to the fit.  
The colors of the filled circles correspond to the 1, 2, and 3 $\sigma$ deviations of a dust mixture from the best fit,
as shown in the legend of Figure~\ref{fig:barchart}.

In the ideal case, the dust samples that correspond with the best fit will also be represented 
in the results of similar dust mixtures.
In the case of GX~5-1 for instance, the best fit consists of two dust samples, leaving two options open, which in the case of the 
best fit does not contribute significantly. This means that out of the 1001 possibilities, there are 91 similar mixtures,
as can be seen in Figure~\ref{fig:barchart} 
by the dominant selection of samples 1 and 7.
When the best fit is unique, we expect a clustering of the similar mixtures around the best fit. This effect can be 
observed in the frame of GX 13+1 in Figure~\ref{fig:barchart} 
for sample 7,  
and to a  lesser degree in GX~5-1 for samples 1 and 7.

If the data are of good quality (i.e., with  high signal-to-noise ratio), which is the case for six out of the nine X-ray binaries,
it becomes possible to observe a preference in the fits for certain dust samples.
This is especially evident in GX 5-1, GX 17+2, and GX 13+1.
When the quality of the data declines, it allows almost every type of dust to be
fitted equally well. This effect can be observed
in the observations of 4U 1702-49, GRS 1758-258, and 4U 1728-34. Therefore, we will not use these sources 
in the discussion of the dust composition.
The implications of Figure~\ref{fig:barchart} 
will be discussed in Section~\ref{dust_comp_gc_ch3}.

\subsection{Silicon abundances and depletion}\label{si_abundance_depletion_ch3}

The silicon K-edge allows the possibility of evaluating silicon in both gas and dust simultaneously. 
Consequently, this allows a study of the abundance and depletion
of silicon on the nine different lines of sight towards the X-ray binaries.  
Table~\ref{table:abundances_depletions} gives the silicon column density ($N^{\mathrm{tot}}$), 
depletion, total silicon abundance ($A_{\mathrm{Si}}$), and 
abundance of silicon in dust ($A_{\mathrm{Si}}^{\mathrm{dust}}$). 
Finally, $A_{\mathrm{Si}}/A_{\sun}$ shows the deviation from the solar abundance of silicon. 
The allowed depletion ranges used in the fits are based upon values from \citet{Jenkins09}. 
The ranges  are given in Table~\ref{table:depletions}.
Since we fit only one edge using dust models, we need to constrain the other elements within reasonable boundaries. 
For the edges for which we do not have dust
features, we use gas absorption-like profiles in the SPEX model. 

\begin{table*}
\caption{Abundances and depletions of silicon} % title of Table
\label{table:abundances_depletions} % is used to refer this table in the text
\centering % used for centering table
\begin{tabular}{l c c c c c} % centered columns (4 columns)
\hline\hline % inserts double horizontal lines
Source& $N^{\mathrm{tot}}_{\mathrm{Si}}$ & depletion & $A_{\mathrm{Si}}$ & $A_{\mathrm{Si}}^{\mathrm{dust}}$ & $A_{\mathrm{Si}}/A_{\sun}$\\ % table heading
& ($10^{18}\,\mathrm{cm}^{-2}$)& & $(10^{-5}\,\mathrm{per}\,\mathrm{H}\,\mathrm{atom})$& $(10^{-5}\,\mathrm{per}\,\mathrm{H}\,\mathrm{atom})$& \\
\hline\noalign{\smallskip} % inserts single horizontal line
GX 5-1 &$1.7\pm0.2$&$0.89\pm0.05$&$2.9\pm0.2$&$2.5\pm0.2$&$0.8^{+0.2}_{-0.1}$\\ % inserting body of the table
GX 13+1 &$1.4\pm0.2$&$0.94\pm0.02$&$4.6\pm0.7$&$4.4\pm0.7$&$1.2\pm0.2$\\ % inserting body of the table
GX 340+00 &$2.9\pm0.6$&$0.77\pm0.10$&$4.5\pm0.9$&$3.6\pm0.9$&$1.2^{+0.2}_{-0.3}$\\ % inserting body of the table
GX 17+2 &$1.1\pm0.2$&$0.95\pm0.03$&$5.2\pm0.4$&$5.1\pm0.4$&$1.3\pm0.1$\\ % inserting body of the table
4U 1705-44 &$0.9\pm0.2$&$0.91\pm0.05$&$4.3\pm1.1$ & $4.0\pm1.1$&$1.1\pm0.3$\\ % inserting body of the table
4U 1630-47 &$4.8\pm1.3$&$0.76^{+0.23}_{-0.28}$&$4.9\pm1.3$&$4.5\pm1.3$&$1.3\pm0.4$\\ % inserting body of the table
4U 1728-34 &$1.4\pm0.4$&$0.95^{+0.02}_{-0.10}$&$4.1\pm1.4$&$3.9\pm1.4$& $1.1\pm0.4$\\ % inserting body of the table
4U 1702-429 &$0.7\pm0.3$&$0.93\pm0.15$&$3.2\pm1.5$&$3.0\pm1.5$&$0.8\pm0.4$\\ % inserting body of the table
GRS 1758-258 &$1.0\pm0.4$&$0.75\pm0.2$&$3.9\pm1.5$ &$2.9\pm1.5$ &$1.0\pm0.4$\\ % inserting body of the table
\hline %inserts single line
average & $1.8\pm0.2$&$0.87\pm0.04$&$4.0\pm0.4$&$3.8\pm0.4$&$1.1\pm0.1$ \\
\hline\hline
\end{tabular}
\tablefoot{Abundances are indicated by $A_{\mathrm{Si}}$. Solar abundances are taken from \citet{Lodders09}.  
$N^{\mathrm{tot}}_{\mathrm{Si}}$ indicates the total column density of silicon (gas and dust).}
\end{table*}

\begin{table}
\caption{Depletion ranges used in the spectral fitting} % title of Table
\label{table:depletions} % is used to refer this table in the text
\centering % used for centering table
\begin{tabular}{l c} % centered columns (4 columns)
\hline\hline % inserts double horizontal lines
Element & depletion range  \\ % table heading
\hline\noalign{\smallskip} % inserts single horizontal line
% \rule{0pt}{2.3ex}
Silicon &$0.41-0.96$  \\ % inserting body of the table
Iron &$0.7-0.97$  \\
Magnesium &$0.47-0.95$\\
Oxygen &$0.02-0.42$  \\
\hline %inserts single line
\end{tabular}
\tablefoot{Depletion ranges in this table are based on depletion values from \citet{Jenkins09}.}
\end{table}

\section{Discussion}\label{discussion_ch3}

\subsection{Dust composition toward the Galactic center}\label{dust_comp_gc_ch3}

The results of the fits of the nine X-ray binaries are summarized in Figure~\ref{fig:barchart}. 
Since this figure 
contains information about the crystallinity, the mineralogy, and the ratio of iron to magnesium,
we discuss each of the properties of the dust separately. We  also focus on the
results of the fits of GX~5-1, GX~13+1, and GX~17+2;  for these sources the quality of the data in terms of 
signal-to-noise ratio around the Si K-edge is the best with respect to the other sources.

\subsubsection{Crystallinity}\label{crystallinity_ch3}

From the fits of all the X-ray binaries we observe that crystalline dust models can be fitted
to the Si K-edge. 
We calculate the ratio of crystalline versus amorphous dust ($\zeta_1$). 
The ratio here is defined as $\zeta_1=$crystalline dust/(crystalline dust + amorphous dust). 
Examining the best fitting dust mixture of GX~5-1, 
we find a value of $\zeta_1=0.12$, and when considering the errors we find an upper limit of  
on the crystallinity of $\zeta_1<0.29$.
GX 13+1 can be analyzed in the same way. Here the obtained ratio of $\zeta_1$ is  $\zeta_1=0.07$,
with  an upper limit of $\zeta_1<0.35$. For GX 17+2 we find the lowest crystallinity value of  $\zeta_1=0.04$ and an upper limit of $\zeta_1<0.17$.
The other sources in the analysis,  focussing on 4U 1630-47, GX 340+00, and 4U 1705-44, show a similar result, 
although the errors on the dust measurements increase because of the data quality.

As seen above and despite the errors, the X-ray binaries with the best signal-to-noise ratios
are best fit by a mixture of mainly amorphous dust and a contribution of crystalline
dust which varies in the range $\zeta_1= 0.04-0.12$.
These amounts of crystalline dust are large 
in comparison with results from the infrared (Section~\ref{sec:introduction}). 
One explanation may be   
that we are observing special lines of sight with freshly produced crystalline dust grains that have not been fully  amorphized by the 
processes in the ISM. This result may be in line with the results from the Stardust mission~\citep{Westphal14}, 
where some of the interstellar silicate dust particles were detected with a crystalline core. 
However, it is unclear why the X-ray lines of sight towards the central Galactic environment would systematically sample
a different environment than the infrared lines of sight. 

An alternative explanation for this apparent discrepancy can be found 
in the different methods used to study the silicate dust. 
XAFSs, especially the features close to the edge, are sensitive to short-range order, whereas 
in the infrared observations are focussed on long-range disorder in the dust particles. 

There are multiple processes to form crystalline and amorphous dust~\citep[e.g.,][]{Dorschner95,Jager03,Speck11}.
The different techniques used in the laboratory to synthesize amorphous dust show that 
some of these samples are glassy, others are porous, and some samples are not homogeneously amorphous,
but show the onset of crystallization. All of these samples produce amorphous infrared dust features, albeit with differences 
in the peak postion of the $10$ and $18\,\mu\mathrm{m}$ features~\citep{Speck11}.
Furthermore, a polycrystalline material can also smear the dust features~\citep{Marra11} in the infrared
and we may thus not perceive sharp crystalline features in the spectrum~\citep{Speck11}.
However, the short range crystalline structure between the atoms in a polycrystal are still intact and XAFS may 
appear in the spectrum. 
Specifically, even if the material becomes slightly amorphous or glassy, XAFS may still appear in the X-ray spectrum, 
but are less pronounced and tend to shift with energy when the material becomes  more disordered~\citep{Mastelaro18}.
Therefore, what may be perceived as amorphous dust in the infrared can still be observed as crystalline dust in the X-rays.
More laboratory research is necessary to make a complete comparison between the crystalline and 
amorphous dust characteristics in the infrared and the X-rays. On the other hand,
high-quality astronomical spectra are necessary to put firmer limits on the amount of crystalline dust 
observed in the spectra of X-ray binaries.

\subsubsection{Iron in silicates}\label{iron_ch3}
Our sample set contains olivines and pyroxenes with different iron and magnesium content. 
Already from the laboratory data it can be observed that the influence of a changing iron content in the dust causes only
small differences in the XAFS (Figure~\ref{fig:all_edges}). 
Small shifts in the peaks can be observed in the laboratory data, 
but in view of the resolution of the \textit{Chandra} spectra, these
subtle changes will not be detected in the spectra.
We note that  samples 2, 3, 4, 5, and 6 almost never contribute to the best fit. 
What these samples have in common is that all of them are iron-poor pyroxenes.
Therefore, there are two reasons why these samples do not fit the edge and the structure of the dust is of greater
influence here than the iron content, since the structural differences in the crystal-types can be distinguished in the models.
Iron is highly depleted in the ISM and is expected to reside in dust particles. Silicates may provide a possibility to store
some of the iron in dust. 
However, even if our fits prefer slightly more iron-rich silicates, not all the iron that should be present in the ISM is accounted for. 
The missing iron can be present in other forms of dust. A possibility could be that iron is included
in metallic form in GEMS~\citep{Bradley94} or in iron nanoparticles~\citep{Bilalbegovi17}.
Our sample set currently contains pyroxene samples with a ratio of Mg/(Fe+Mg)=0.6. This is very similar to the  \citet{Kemper04} value of Mg/(Fe+Mg)=0.5.
Previous studies do not show  strong evidence that this ratio should be much lower. 
Further constraints on the iron content would be provided by a multiple-edge fitting.
The magnesium and the iron K-edge will give a 
direct impression of magnesium and iron in dust. Depending on the brightness of the source, the column density along the line 
of sight and the telescope and instruments that are used, it is possible to observe these edges. 
In the case of the Fe K-edge, the current instruments 
are not sensitive enough around the edge to detect the XAFS, but the future observatory \textit{Athena}
will be able to observe the Fe K-edge in detail~\citep{Rogantini18}. 

\subsubsection{Olivines, pyroxenes, and oxides}\label{oli_pyr_ox_ch3}
The difference between iron-rich and iron-poor dust models in the laboratory data is subtle; instead,  
the difference between olivines, pyroxenes, and 
quartz types in crystalline form is striking. This means that it is easier to identify differences in the mineralogy.
In general, we observe that in almost all of the X-ray binaries, the best fitting dust mixture includes an olivine dust type 
(whether amorphous or crystalline), but not all the data have similar signal-to-noise ratios.
The three X-ray binaries with the best signal-to-noise ratios, namely GX~5-1, 
GX~17+2, and GX~13+1, show a preference for amorphous olivine  in the best fit and in the fits within 1$\sigma$. 
The ratio of olivine
to pyroxene can be expressed as $\zeta_2=$olivine/(pyroxene+olivine). For GX~5-1, 
GX~17+2, and GX~13+1 $\zeta_2=1$ in case of the best fits. However, within 1$\sigma$ from the best fitting dust mixtures 
it is possible to obtain lower values of $\zeta_2$ with 
a minimum of $\zeta_2=0.8$, meaning that we can obtain a good fit with a maximum of 20\% pyroxene in the dust mix. 

Thus, in the central Galactic environment we do not find much variation in the best fitting dust mixture. 
We compare this result with studies of silicates in the infrared. 
By analyzing the $10\,\mu\mathrm{m}$ silicate feature, 
\citet{Kemper04} also find that olivine glass accounts for most of the silicate mass in the diffuse ISM along the line of 
sight toward the GC.
However, in the infrared, variations in the stoichiometry of the dust have been found along different lines of sight.
Fitting both the $10\,\mu\mathrm{m}$ and $18\,\mu\mathrm{m}$ silicate features of Wolf-Rayet stars representing both the local ISM
and the GC,
\citet{Chiar06} find that a mix of olivine and pyroxene silicates produces a good match to their data, and that a greater 
contribution by mass of pyroxene dust is required.
Observing the same line of sight as~\citet{Kemper04},
~\citet{Min07} find a stoichiometry of the silicate dust that lies  between that of olivine and
pyroxene.
In future X-ray studies it will therefore be interesting to investigate samples with a stoichiometry 
 between that of olivine and pyroxene. 
 
The role of $\mathrm{SiO}_2$ dust in the ISM is not well known. 
This type of dust may form in the ISM and may be present in the form of 
$\mathrm{SiO}_2$ nanoparticles, although there is limited insight into how these dust particles 
may form and they have not been
detected in the ISM~\citep{Li02,Krasnokutski14}. The presence of $\mathrm{SiO}_2$ may be supportive of the formation
of grains in the interstellar medium~\citep[][and references therein]{Krasnokutski14}.
The three $\mathrm{SiO}_2$ samples in our sample set  
can be fitted within one sigma of the best fit, most notably in 4U 1630-47. Considering the overall contribution of 
$\mathrm{SiO}_2$ in the fits of GX~13+1, GX~5-1 and GX 17+2,
we do not find evidence that  $\mathrm{SiO}_2$ is  the dominant component in interstellar dust. 

\subsection{Silicon abundances and depletion}

The results from Table~\ref{table:abundances_depletions} allow us to study the dense environment in the Galactic plane and in the 
vicinity of the GC. 
The abundance of silicon can be derived from infrared data, using observations of the 10 and 18 $\mu$m features~\citep{Aitken84,Roche85,Tielens96}.  
In the local solar environment the silicon abundance in dust can be derived as $5.2\pm1.8\times10^{-5}$ per H-atom, 
using data from~\citet{Roche84} and \citet{Tielens96} who observed nearby Wolf-Rayet stars. 
On sight lines toward the GC the abundance of silicon in dust is often found to be lower, 
namely $3.0\pm1.8\times10^{-5}$ per H-atom~\citep{Roche85}.
This discrepancy may be caused by the presence of large particles ($>3\,\mu$m) near the GC.
The results of the dust abundance near the GC are also more uncertain, since these infrared silicon abundances
depend on an estimate of the visual extinction ($A_V$) derived from the  $N_\mathrm{H}$-to-$A_V$ ratio of the local solar neighborhood~\citep{Bohlin78}
and this method may be more uncertain toward the GC~\citep{Tielens96}.
On average our results of the silicate dust abundance, $4.2\pm0.4\times10^{-5}$ per H-atom 
fall  between the abundances found by~\citet{Roche85} to the GC and~\citet{Roche84,Tielens96} in the local 
solar environment, 
but individual lines of sight deviate significantly from the average. 
We did not find a relation between the silicon dust abundance and the distance of the source from the GC 
($\mathrm{R_{GC}}$), also called Galactocentric radius. There also appears  to be no obvious relation between the abundance of silicon
and the distance from the plane of the Galaxy, but this can be attributed to the proximity of all our sources to the plane.

\begin{figure}
  \begin{center}
  \includegraphics[scale=0.5]{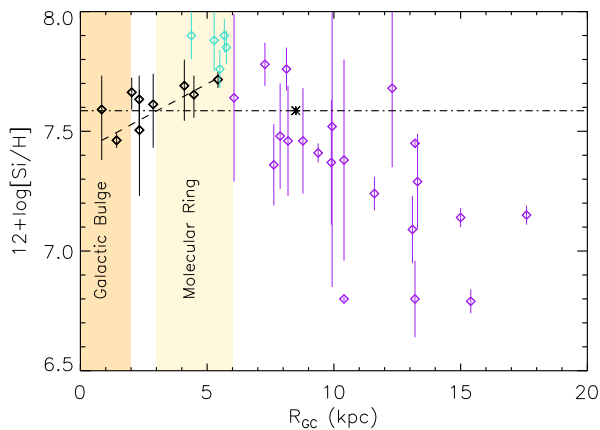}
  \caption{\small{Total silicon abundance expressed in logarithmic units, 
  with hydrogen by definition 12.0 vs. the Galactocentric distance in kpc. The star 
  indicates the position of  the Sun and the two yellow bands indicate the position of the Galactic bulge and the molecular ring.
  The dash-dotted line indicates the solar abundance and the dashed line the gradient in the abundances obtained 
  from the nine X-ray  binaries.}}
  \label{fig:relation_tot_ab}
   \end{center} 
 \end{figure}

In addition to  the abundance of silicon in dust, we can also investigate the total abundance of silicon along the nine lines of sight
toward the X-ray binaries. For elements such as Fe, Mg, O, and Si, an increase in the abundance 
toward the GC is observed~\citep{Pedicelli09,Rolleston00,Davies09}. The proximity  of the X-ray binaries to the GC
gives us a unique opportunity to study the behavior of this gradient in the inner region of the Galaxy, 
 around 0.9 kpc from the center of the Galaxy.
~\citet{Rolleston00} found  a Galactic abundance gradient 
of silicon of $-0.06\pm0.01$ dex/kpc for Galactocentric radii $\mathrm{R_{GC}}>6$ kpc. 
For Galactocentric radii $\mathrm{R_{GC}}<6$ kpc  no clear gradient was found, as was shown by~\citet{Davies09}. They find 
a gradient for the azimuthal abundance of $-0.8\pm0.1$ dex/kpc for silicon, but the observations have a large scatter. 
Of  course the area of the Galactic bulge provides a different environment from the surrounding Galactic disk.
Using Figure~\ref{fig:relation_tot_ab}, 
we can investigate a possible gradient of abundances toward the Galactic plane. 
Here we show the total abundance of silicon versus $\mathrm{R_{GC}}$.
The abundances obtained from the X-ray binaries are shown in black. 
The other data points are abundances of silicon of a sample of B-type stars (in purple) from~\citet{Rolleston00} 
and abundances derived from observations of cepheids (in light blue) from~\citet{Andrievsky02} for comparison.
The star symbol at 8.5 kpc indicates the position of the sun and the two yellow bands show the environments of the Galactic bulge 
and the Galactic molecular ring. 
The dash-dotted line shows the solar abundance of silicon. 
We find a gradient of $0.06\pm0.01$ dex/kpc for the abundances derived from the X-ray binary observations. 
The gradient is indicated by the dashed line.
Instead of the increase in the silicon abundance observed at radii larger than 6 kpc from the center, we observe a slight decrease.
We note that the error on this gradient is large, and  Figure~\ref{fig:relation_tot_ab}
shows that all the abundances are close to solar.
If this decrease is real, it might be caused by an increase in the typical grain size of the silicate grains in the 
Galactic central area to which the X-rays are not sensitive. In that case the X-rays are a probe for the volume of large grains. 
This is supported by the observation of a scattering feature just before the edge indicating the possible presence of 
large grains, as described in Z17.\footnote{\tiny{Future studies will explore different grain size distributions, as well as the implementation of non-spherical 
grains to analyze the scattering feature. 
However, we can  already predict some of the effects of the presence of non-spherical grains on our analysis. 
Non-spherical grains mainly affect the scattering features at the edges of the spectra. 
The scattering feature just before the edge is the most prominent scattering feature. 
Therefore, when we use non-spherical elongated grains, we  observe the most significant changes in this feature. 
Examples of how this feature changes for oblate and prolate grains are shown in~\citet{Hoffman16}, figure 9.
When we compare these grains to spherical grains we can observe that the changes in the XAFS features are modest when the shape of the grain changes.
This is expected since the spherical grains represent an average radius of the major and minor axis of the prolate and oblate grains.}} 
However, we  note here that the error on the distance should also be taken into account 
before a firm conclusion about the gradient can be made. Furthermore, the errors on the abundance measurements can be reduced by
more and longer observations of the sources. In the case of GX~13+1 the errors are already small due to the number of observations 
we included in the fit, and therefore serves as a good example to show the benefit of new observations.
All these elements considered, we can conclude that the increase in abundance of silicon at Galactocentric
radii $>6$ kpc is not observed in the inner part of the Galaxy.

\begin{figure}
  \begin{center}
   \includegraphics[scale=0.5]{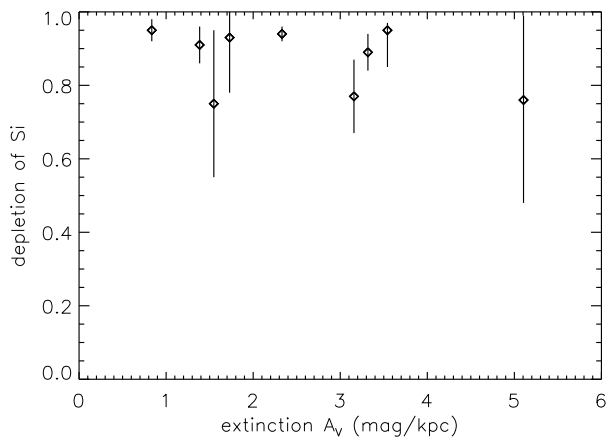}
  \caption{\small{Depletion vs. visual extinction ($A_V$).}}
  \label{fig:relation_depletion}
   \end{center} 
 \end{figure}

The depletion of elements such as Fe, Mg, O, and Si 
in to dust shows a correlation with the extinction along the line of sight~\citep{Jenkins09, Voshchinnikov10}. 
Where~\citet{Voshchinnikov10} probe the depletion up to a distance of 7 kpc from the Sun, we are able to observe the depletion of silicon at larger distances and in the less  explored environment of the
Galactic central region. 
Furthermore, we are able to observe both gas and dust simultaneously, and in this way we
obtain a direct measure of the depletion. 
Assuming the relation for visual extinction $N_{H}/A_{V} = 1.9\times10^{21} \mathrm{cm}^{-2} \mathrm{mag}^{-1}$~\citep{Bohlin78}, 
is still valid in the dense environment of the central part of the Galaxy,
we obtain the result of the depletions versus the extinction, $A_{V}$, shown in Figure~\ref{fig:relation_depletion}. 
Not all of the X-ray binaries follow the average extinction of 2 mag/kpc. Four lines of sight  have extinctions higher
than 3 mag/kpc. We do not observe a clear trend in the data. 
This lack of correlation can be explained by the environment in which the X-ray binaries reside.
The central part of the Galaxy has a complex structure as we already noted earlier in this section. All the X-ray binaries 
are located in this area, so we may be observing local variations in the ISM. 

\section{Summary and conclusion}\label{summ_concl_ch3}
In this paper, we fitted the Si edge of of the absorbed spectra along nine different lines of sight.
We used a total of 14 new dust extinction profiles,
representing to a good degree the silicate content of interstellar dust.
We measured the absorption profiles of these 14 interstellar dust analogs at
the Soleil synchrotron facility in Paris. 
The laboratory absorption measurements were converted into extinction cross sections
in order to obtain models  suitable for interstellar dust studies.
We obtained the following results: 
\begin{itemize}
 \item We find that most lines of sight can be  fitted well by amorphous olivine. 
 The contribution of crystalline dust to the fits is larger than found in the infrared.
 For the sources with the best signal-to-noise ratios (i.e., GX~5-1, GX~13+1, and GX~17+2), 
 we find values of the crystallinity in the range  $\zeta_1=0.04-0.12$, 
 with upper limits on these values of $\zeta_1$  between 0.17 and 0.35.
 A possible explanation may lie in the nature of X-rays, which is such that it facilitates the  study of the short-range order 
 between the atoms, contrary to the long-range disorder in the infrared. 
 In this way, we may observe crystallinity in polycrystalline and partly glassy material.
 More high-quality observations will allow us to put further constraints on this parameter.
\item Iron-poor pyroxenes are not preferred in the fits. It is  difficult, however,  to put a precise limit on the amount of iron 
in silicates, since the Si K-edge is not very sensitive to changes in the iron content. In order to investigate the contribution of 
iron in silicates we need to involve the Fe K-edge. Observations of the Fe K-edge in X-ray binaries will be possible with the 
future \textit{Athena} observatory. 
\item In almost all of the X-ray binaries, the dust mixture best fitting the Si K-edge includes an olivine dust type. 
\item For the first time, we study the GC environment in the X-rays using the Si K-edge in the spectra of X-ray binaries
located in this environment at $\mathrm{R_{GC}} < 4$ kpc. 
For every line of sight, we obtained the total abundance of silicon, the dust abundance, and the depletion.
We investigated trends 
for the total abundance versus the Galactocentric distance. 
We observe a decrease in the silicon abundance toward the GC with a gradient of $0.06\pm0.01$
dex/kpc.  This may be caused by 
silicon atoms locked up in large ($>3\,\mu$m) dust particles in these dense environments. 
\end{itemize} 

\begin{acknowledgements}
We would like to thank the anonymous referee for providing us with helpful comments.
Dust studies at Leiden Observatory are supported
through the Spinoza Premie of the Dutch science agency, NWO. E.C. and D.R. acknowledge support from NWO-Vidi
grant 639.042.525.
H.M. and P.M. are grateful for the support of the Deutsche Forschungsgemeinschaft under Mu 1164/8-2 and Mu 1164/9-1.
We acknowledge SOLEIL for provision of
synchrotron radiation facilities, and we would like to thank Delphine Vantelon for
assistance in using beamline LUCIA. 
This research made use of the Chandra Transmission
Grating Catalog and archive (http://tgcat.mit.edu). We also made use of the FLUO self-absorption correction code provided by 
Daniel Haskel.
\end{acknowledgements}

\begin{appendix}

\section{Broadband spectral fits of the individual sources}\label{appendix:data_tables_ch3}
\begin{figure}
 \begin{center}
 \includegraphics[scale=0.5]{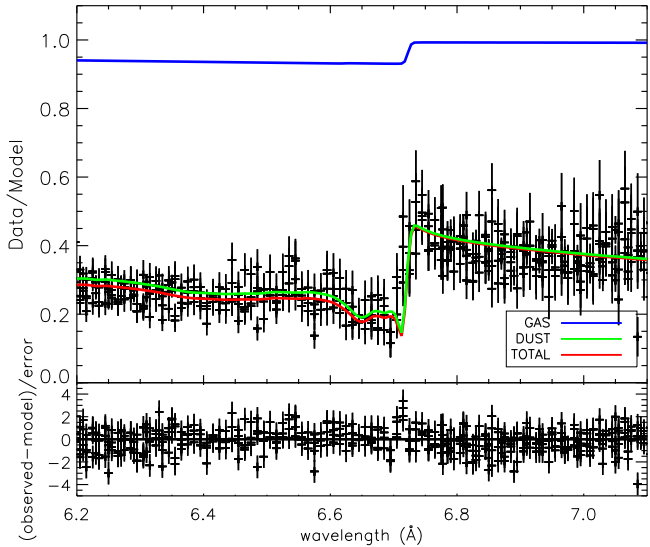}
 \caption{\small{Si K-edge of 4U 1630-47.}}
 \label{fig:4U 1630-47}
  \end{center} 
\end{figure}

\begin{figure}
 \begin{center}
 \includegraphics[scale=0.5]{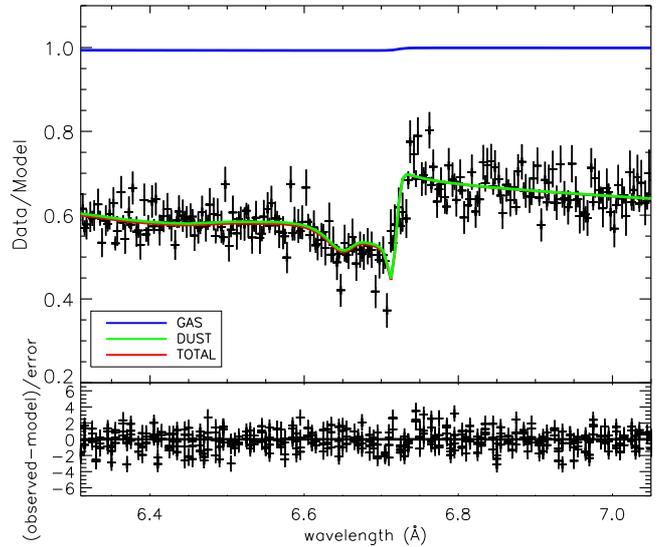}
 \caption{\small{Si K-edge GX 13+1.}}
 \label{fig:GX_13+1}
  \end{center} 
\end{figure}

We give a detailed overview of the data obtained from the best fits of X-ray binaries GX 5-1, 4U 1630-47, GX 13+1,
4U 1702-429, 4U 1728-34, GX 340+00, GRS 1758-258, GX 17+2, and 4U 1705-44. In the tables of this section we give the parameter values corresponding 
to the best fits of each of the sources. The best fits of each of these sources are shown in the corresponding figures.\\
\textit{GX 5-1}:
The fitting of GX 5-1 is explained in Section~\ref{model_proc_ch3} for illustration.
We found a column density of $5.8\pm0.2\times10^{22}\mathrm{cm}^{-2}$.
This value can be compared to previous studies: the column density of GX~5-1 was measured by \citet{Predehl95} to range 
between 2.78 and $3.48\times10^{22}\,\mathrm{cm}^{-2}$ depending on the continuum model. 
More recent values of the column density are $2.8\times10^{22}\,\mathrm{cm}^{-2}$
with \textit{Chandra} data by \citet{Ueda05} and $3.07\pm0.04\times10^{22}\,\mathrm{cm}^{-2}$ by \citet[][using ASCA archival data]{Asai00}.
In Z17 we made use of the short (0.24 ks) 
observation of obsid 716 in order to minimize the effect of pile-up on the estimate of the column density. 
This resulted in a column density of $3.4\pm0.1\times10^{22}\,\mathrm{cm}^{-2}$. 
These values are lower than the one we find for the spectra used in this study.
However, there is a considerable time difference between the observation used in this analysis (July 2017) 
and the previous observation in TE mode by Chandra (July 2000). 
From the observation listed above, we already noted that the observed column density of the source can vary. 
This variation may be associated with changes over time intrinsic to the source. 
Such differences in the column density can also be observed  in EXO 0748-676, for example~\citep{Peet09}. 
Furthermore, GX 5-1 deviates from the linear relation between the scattering optical depth 
and the column density~\citep[see Figure 7 in ][]{Predehl95}. 
When the interstellar medium is solely responsible for the total amount of absorption, 
a lower column density is expected for GX 5-1 with respect to the observed scattering optical depth. 
Since $N_\mathrm{H}$ is observed to be larger, the increase can be associated with the source.\\ 
\textit{4U 1630-47}:
There are four data sets  used in the fitting of the
4U 1630-47 (Table~\ref{table:4U163047}, Figure~\ref{fig:4U 1630-47}).
The source continuum  is modeled using two blackbody models. All observations 
show outflowing gas, which is modeled by the XABS model. 
The ionization parameter $\xi$ in the XABS model is defined as $\xi = L/nr^2$, where $L$ is the ionizing luminosity, 
$n$ the gas density, and $r$ the distance of the gas from the source.
We find a column density of $N_\mathrm{H}=9.7\pm0.1\times10^{22}\mathrm{cm}^{-2}$,
making it the densest line of sight in our study. This value is in agreement with ~\citet{Neilsen14}, among others, 
who find a best fit column density of $N_\mathrm{H}= 9.4^{+0.5}_{-1.1}\times10^{22}\mathrm{cm}^{-2}$.\\
\textit{GX~13+1}:
The continuum of GX~13+1 is fitted with a disk blackbody and Comptonization model
(Table~\ref{table:GX131}, Figure~\ref{fig:GX_13+1}). 
All the observations have a XABS component in order to 
model outflowing gas from the source. In the case of 
the observation with obsid 11814, a second XABS model is introduced in order to fit 
the non-outflowing ionized gas along the line of sight. The column density of GX 13+1 is in agreement with values found by
~\citet{Pintore14} and~\citet{Dai14}.

\begin{figure}
 \begin{center}
 \includegraphics[scale=0.5]{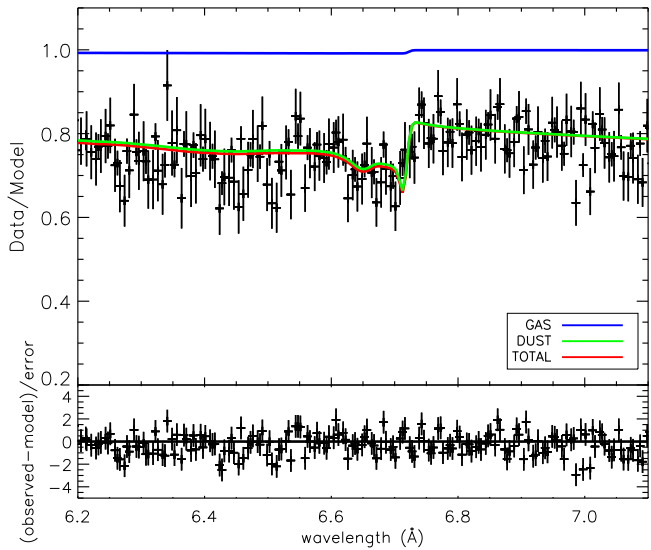}
 \caption{\small{Si K-edge 4U 1702-429.}}
 \label{fig:4U_1702-429}
  \end{center} 
\end{figure}

\begin{figure}
  \begin{center}
  \includegraphics[scale=0.5]{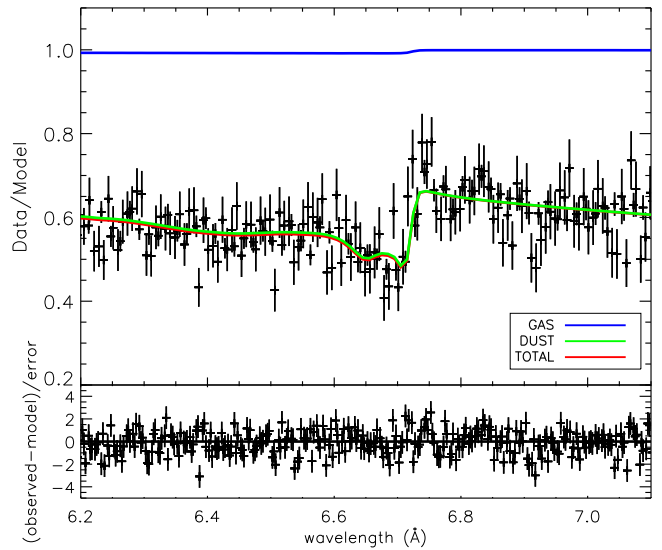}
  \caption{\small{Si K-edge 4U 1728-34.}}
  \label{fig:4U_1728-34}
   \end{center} 
 \end{figure}

\textit{4U~1702-429}: This X-ray binary was modeled using a disk blackbody and a Comptonization model, resulting in a value of $N_\mathrm{H}$
$2.3\pm0.2\times10^{22}\mathrm{cm}^{-2}$, similar to the results found by~\citet{Iaria16}
(Table~\ref{table:4U1702429_4U172834_GX34000}, Figure~\ref{fig:4U_1702-429}
).\\
\textit{4U 1728-34}: In the case of 4U 1728-34 the continuum was also modeled by using a disk blackbody and a Comptonization model
(Table~\ref{table:4U1702429_4U172834_GX34000}, Figure~\ref{fig:4U_1728-34}
). 
This source is a bursting low-mass X-ray binary, and two bursts occur in the spectrum of obsid 2748. These bursts do not affect
the modeling of the Si K-edge.
The $N_\mathrm{H}$ is in agreement with values found by~\citet{Salvo00} of $3.1\pm0.1\times10^{22}\mathrm{cm}^{-2}$, 
using a similar modeling and data from the
\textit{BeppoSAX} satellite, which covers a wide energy range of 0.12-100 keV, making it especially suitable for hydrogen
column density measurements. \\
\textit{GX 340+00}: is  fitted well by a power-law model in combination with a blackbody model. We added a Gaussian model  to fit 
the iron $\mathrm{K}_{\alpha}$ emission-line feature around 6.4 keV. The obtained column density of hydrogen is $6.6\pm0.2\times10^{22}\,\mathrm{cm}^{-2}$. 
This value is similar to results from~\citet{Seifina13}, who make use of data from \textit{BeppoSAX} and \textit{RXTE} 
and find values of $\sim5.5-6.5\times10^{22}\,\mathrm{cm}^{-2}$ 
for several models. It is below values found by~\citet{Cackett10} of $0.9-1.1\times10^{23}\,\mathrm{cm}^{-2}$ using XMM-\textit{Newton} data.

%Table~\ref{table:GRS1758258_GX172_4U170544} shows the results for GRS 1758-258, GX 17+2 and 4U 1705-44. 
\textit{GRS 1758-258}:
The continuum of GRS 1758-258 is best fit by a blackbody and a steep power 
law function~\citep{Smith01}
(Table~\ref{table:GRS1758258_GX172_4U170544}, Figure~\ref{fig:GRS_1758-258}). \\
\textit{GX 17+2}: The continuum of GX 17+2 modeled using a blackbody model in combination with 
power-law function (Table~\ref{table:GRS1758258_GX172_4U170544}, Figure~\ref{fig:GX_17+2}). 
The column density is estimated by \citet{Cackett09} to lie between $3.12\pm0.05$ and $4.63\pm0.08\times10^{22}\,\mathrm{cm}^{-2}$,
using \textit{Chandra} data. This column density is higher than the value found in our analysis of $2.0\pm0.1\times10^{22}\,\mathrm{cm}^{-2}$. 
However, our analysis is in agreement with the analysis of \citet{Salvo200}, who use data from~\textit{BeppoSAX}. \\
\textit{4U 1705-44}: The continuum of 4U 1705-44 is also modeled using a blackbody model in combination with 
power-law function (Table~\ref{table:GRS1758258_GX172_4U170544}, Figure~\ref{fig:4U_1705-44}).
The column density is in agreement with data from \textit{BeppoSAX}~\citep[][model 2]{Piraino16}.

\begin{table*}
\caption{Best fit parameters for GX 5-1} % title of Table
\label{table:gx51} % is used to refer this table in the text
\centering % used for centering table
\begin{tabular}{l c c } % centered columns (4 columns)
\hline\hline % inserts double horizontal lines
\rule{0pt}{2.3ex}
Obsid  & 13714 & 13715  \\ \hline  \noalign{\smallskip}% table heading
$N_{\mathrm{H}}^{\mathrm{cold}}$ ($10^{22}\,\mathrm{cm}^{-2}$)& \multicolumn{2}{c}{$5.8\pm0.2$}\\
$k_\mathrm{B}\mathrm{T}_{\mathrm{bb}}$ (keV) &$0.31\pm0.01$&$0.33\pm0.01$ \\
$k_\mathrm{B}\mathrm{T}_{\mathrm{0\,comt}}$ (keV) &$0.30\pm0.03$&$0.30\pm0.03$ \\
$k_\mathrm{B}\mathrm{T}_{\mathrm{1\,comt}}$ (keV) &$28\pm9$&$23^{+12}_{-6}$\\
$\tau_{\mathrm{comt}}$ (keV) &$1.0^{+0.2}_{-0.7}$&$1.2\pm0.6$\\
$F_{0.5-2\,\mathrm{keV}}\,(10^{-10}\,\mathrm{erg}\,\mathrm{cm}^{-2}\mathrm{s}^{-1})$&$3.8\pm0.4$&$3.6\pm0.4$\\
$F_{2-10\,\mathrm{keV}}\,(10^{-8}\,\mathrm{erg}\,\mathrm{cm}^{-2}\mathrm{s}^{-1})$&$1.8\pm0.2$&$1.7\pm0.2$\\  \cline{2-3}
$C^2/\nu$& \multicolumn{2}{c}{2026/1603}  \\ 
\hline %inserts single line
\end{tabular}
 \tablefoot{This fit was produced using the following SPEX models: a blackbody model, a Comptonization model, 
 the AMOL model, cold gas model (HOT with $k_\mathrm{B}T=5\times10^{-4}$), and XABS.}
\end{table*}

%4U 1630-47
 
\begin{table*}
\caption{Best fit parameters for 4U 1630-47} % title of Table
\label{table:4U163047} % is used to refer this table in the text
\centering % used for centering table
\begin{tabular}{l c c c c} % centered columns (4 columns)
\hline\hline % inserts double horizontal lines
\rule{0pt}{2.3ex}
Obsid  & 13714 & 13715 & 13716 & 13717 \\ \hline  \noalign{\smallskip}% table heading
$N_{\mathrm{H}}^{\mathrm{cold}}$ ($10^{22}\,\mathrm{cm}^{-2}$)  & \multicolumn{4}{c}{$9.7\pm0.1$}\\
$k_\mathrm{B}T_{\mathrm{bb1}}$ (keV) &$0.62\pm0.01$&$0.62\pm0.01$ &$0.60\pm0.01$&$0.63\pm0.01$\\ 
$k_\mathrm{B}T_{\mathrm{bb2}}$ (keV) &$1.2\pm0.3$ & $1.2\pm0.4$& $1.2\pm0.4$&$1.2\pm0.5$ \\
$N_\mathrm{H}^{\mathrm{xabs}} (10^{22} \mathrm{cm}^{-2})$& \multicolumn{4}{c}{$9.2\pm0.2$}\\
$\log\xi^{\mathrm{xabs}}\,(\mathrm{erg}\,\mathrm{cm}\,\mathrm{s}^{-1})$ & \multicolumn{4}{c}{$4.1\pm0.2$}\\
$zv_{\mathrm{out}}^{\mathrm{xabs}}$ ($10^2\,\mathrm{kms}^{-1}$) & \multicolumn{4}{c}{$-1.0_{-1.0}^{+1.3}$}\\ 
$N_\mathrm{H}^{\mathrm{xabs2}} (10^{22} \mathrm{cm}^{-2})$& \multicolumn{4}{c}{$9.0\pm0.2$}\\
$\log\xi^{\mathrm{xabs2}}\,(\mathrm{erg}\,\mathrm{cm}\,\mathrm{s}^{-1})$ & \multicolumn{4}{c}{$4.3\pm0.1$}\\
$zv_{\mathrm{out}}^{\mathrm{xabs2}}$ ($10^2\,\mathrm{kms}^{-1}$) & \multicolumn{4}{c}{$-7.8\pm0.3$}\\ 
$F_{0.5-2\,\mathrm{keV}}\,(10^{-12}\,\mathrm{erg}\,\mathrm{cm}^{-2}\mathrm{s}^{-1})$& $8.2\pm2$&$8.2\pm2.0$& $7.8\pm1.9$&$8.0\pm1.8$\\
$F_{2-10\,\mathrm{keV}}\,(10^{-9}\,\mathrm{erg}\,\mathrm{cm}^{-2}\mathrm{s}^{-1})$&$4.1\pm0.9$&$4.0\pm0.9$&$3.8\pm0.8$&$4.4\pm0.9$\\  \cline{2-5}
$C^2/\nu$& \multicolumn{4}{c}{5258/4028}  \\ 
\hline %inserts single line
\end{tabular}
\tablefoot{This fit was produced using the following SPEX models: two blackbody models, 
the AMOL model, the cold gas model (HOT with $k_\mathrm{B}T=5\times10^{-4}$), and two XABS models. Both XABS models are coupled
to  the four observations, since these observations were performed in succession.}
\end{table*}

%GX 13+1

\begin{table*}
\caption{Best fit parameters for GX 13+1} % title of Table
\label{table:GX131} % is used to refer this table in the text
\centering % used for centering table
\begin{tabular}{l c c c c} % centered columns (4 columns)
\hline\hline % inserts double horizontal lines
\rule{0pt}{2.3ex}
Obsid  & 11814 & 11815 & 11816 & 11817 \\ \hline  \noalign{\smallskip}% table heading
$N_{\mathrm{H}}^{\mathrm{cold}}$ ($10^{22}\mathrm{cm}^{-2}$)  & \multicolumn{4}{c}{$3.1\pm0.1$}\\ %nh  :   3.58160E-02 Errors:  -5.00340E-03 ,   2.91441E-04
$k_\mathrm{B}T_{\mathrm{dbb}}$ (keV) &$0.72\pm0.18$&$0.99\pm0.06$&$0.87\pm0.25$ & $0.32\pm0.14$\\ 
$k_\mathrm{B}T_{\mathrm{0\,comt}}$ (keV) &$0.70\pm0.03$&$0.79\pm0.03$ &$0.80\pm0.06$&$0.70\pm0.02$\\
$k_\mathrm{B}T_{\mathrm{1\,comt}}$ (keV) & $12\pm3$&$11\pm2$&$12^{+8}_{-3}$&$11\pm1$\\
$\tau_{\mathrm{comt}}$ (keV) &$1.4\pm0.2$&$0.54\pm0.09$& $0.20^{+0.30}_{-0.19}$&$1.9^{+0.4}_{-0.6}$\\
$N_{\mathrm{H}}^{\mathrm{xabs}_1}$ ($10^{23}\,\mathrm{cm}^{-2}$)&$1.1\pm0.1$&$1.8\pm0.1$&$4.1\pm0.1$&$3.2\pm0.1$\\
$\log\xi^{\mathrm{xabs}_1}$& $4.3\pm0.5$ & $4.3\pm0.3$&$4.3\pm0.1$&$4.5\pm0.1$ \\
$zv_\mathrm{out}^{\mathrm{xabs}_1}$($10^2$ $\mathrm{kms}^{-1}$) &$-4.4^{+3.0}_{-5.3}$&$-5.5^{+1.9}_{-1.5}$&$-6.2^{+3.7}_{-2.1}$&$-3.5\pm1.4$\\ 
$N_{\mathrm{H}}^{\mathrm{xabs}_2}$ ($10^{21}\,\mathrm{cm}^{-2}$)&$1.1\pm0.2$& - & - & - \\
$\log\xi^{\mathrm{xabs}_2}\,(\mathrm{erg}\,\mathrm{cm}\,\mathrm{s}^{-1})$&$3.0\pm0.1$& - & - & - \\
$zv^{\mathrm{xabs}_2}$($10^2\,\mathrm{kms}^{-1}$) & $>-0.1$ & - & - & - \\ 
$F_{0.5-2\,\mathrm{keV}}\,(10^{-10}\,\mathrm{erg}\,\mathrm{cm}^{-2}\mathrm{s}^{-1})$&$1.4\pm0.1$&$1.8\pm0.2$&$1.8\pm0.2$&$1.6\pm0.2$ \\
$F_{2-10\,\mathrm{keV}}\,(10^{-9}\,\mathrm{erg}\,\mathrm{cm}^{-2}\mathrm{s}^{-1})$&$5.6\pm0.6$&$6.6\pm0.7$&$6.5\pm0.7$&$6.8\pm0.7$\\  \cline{2-5}
$C^2/\nu$& \multicolumn{4}{c}{5910/4738}  \\ 
\hline %inserts single line
\end{tabular}
\tablefoot{This fit was produced using the following SPEX model components: a disk blackbody, a Comptonization model, 
the AMOL model, the cold gas model (i.e., HOT with $k_\mathrm{B}T=5\times10^{-4}$), and two XABS models. }
\end{table*}

%4U 1702-429, 4U 1728-34 and GX 340+00

 \begin{figure}
  \begin{center}
  \includegraphics[scale=0.5]{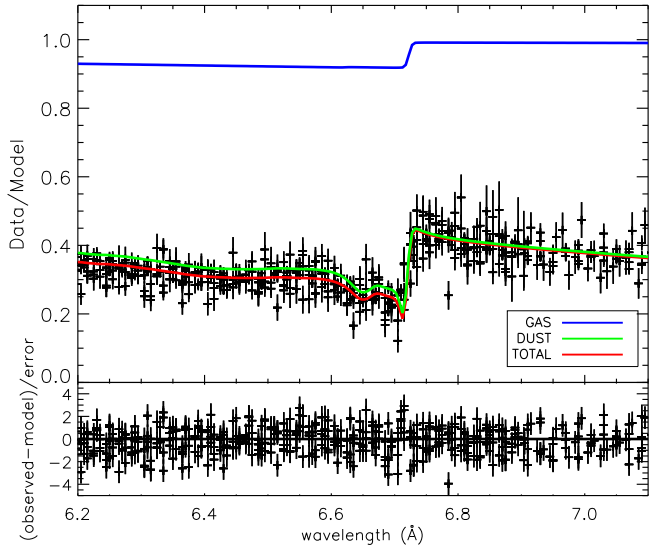}
  \caption{\small{Si K-edge GX 340+00.}}
  \label{fig:GX_340+00}
   \end{center} 
 \end{figure}

\begin{table*}
\caption{Best fit parameters for 4U 1702-429, 4U 1728-34, and GX 340+00} % title of Table
\label{table:4U1702429_4U172834_GX34000} % is used to refer this table in the text
\centering % used for centering table
\begin{tabular}{l c c c c c c} % centered columns (4 columns)
\hline\hline % inserts double horizontal lines
\rule{0pt}{2.3ex}
Source    &  4U 1702-429 & 4U 1728-34 & \multicolumn{4}{c}{GX 340+00}\\
Obsid  & 11045 & 2748 & 1921 & 18085 & 19450 & 20099\\ \hline  \noalign{\smallskip}% table heading
$N_{\mathrm{H}}^{\mathrm{cold}}$ ($10^{22}\,\mathrm{cm}^{-2}$)  & $2.3\pm0.2$ & $3.5^{+0.2}_{-0.5}$ & \multicolumn{4}{c}{$6.6\pm0.2$}\\
$k_\mathrm{B}T_{\mathrm{bb}}$ (keV) & - & - & $2.1^{+0.7}_{-0.1}$ & $1.1\pm0.1$ & $1.0\pm0.1$ & $1.0\pm0.1$ \\  
$\Gamma_{\mathrm{pow}}$ & - & - & $1.6\pm0.1$ & $1.5\pm0.1$ & $2.0\pm0.1$ & $1.8\pm0.1$\\
$k_\mathrm{B}T_{\mathrm{dbb}}$ (keV) &  $0.74^{+0.22}_{-0.10}$ &  $0.27^{+0.01}_{-0.05}$ & - & - & - & - \\  %0.73943     Errors:  -0.10219     ,   0.11288\\ t   :   0.27477     Errors:  -4.77468E-02 ,   9.01854E-03
$E_{\mathrm{gaus}}$ (keV) & - & - & $6.6\pm0.1$ & - & $6.4\pm0.1$ &  $6.3\pm0.1$ \\
$FWHM_{\mathrm{gaus}}$ (keV) & - & - & $0.40\pm0.08$ & - & $2.3^{+0.2}_{-0.1}$ & $2.9\pm0.2$ \\
$k_\mathrm{B}T_{0\,\mathrm{compt}}$ &  $0.47^{+0.21}_{-0.07}$ & $0.44^{+0.05}_{-0.10}$ & - & - & - & -\\ %  t0  :   0.43827     Errors:  -0.10272     ,   4.52996E-02
$k_\mathrm{B}T_{1\,\mathrm{compt}}$ (keV) &  $2.5^{+0.3}_{-0.5}$ & $21^{+15}_{-17}$ & - & - & - & - \\ % t1  :    21.065     Errors:   -17.563     ,    14.779  
$\tau_\mathrm{compt}$ &  $7.0^{+7.3}_{-0.6}$ & $2.5^{+3.4}_{-2.1} $ & - & - & - & -\\
$F_{0.5-2\,\mathrm{keV}} \,(10^{-11}\mathrm{erg}\, \mathrm{cm}^{-2} \mathrm{s}^{-1})$&$5.6\pm1.0$&$0.8\pm0.1$& $2.5\pm0.2$&$4.0\pm0.4$&$4.3\pm0.3$ &$4.4\pm0.3$\\
$F_{2-10\,\mathrm{keV}} \,(10^{-9}\mathrm{erg}\, \mathrm{cm}^{-2} \mathrm{s}^{-1})$&$0.6\pm0.1$&$1.7\pm0.1$&$6.2\pm0.4$&$10\pm1$&$7.6\pm0.5$& $9.4\pm0.7$\\  \cline{4-7}%Parameter   1    2 norm:    3.1933     Errors:  -1.04055E-02 ,   1.04406E-02
$C^2/\nu$&  2350/2144 & 1403/1326 & \multicolumn{4}{c}{4777/3954} \\ 
\hline %inserts single line
\end{tabular}
\tablefoot{The fit of 4U 1702-429 was produced using the following SPEX model components:
a disk blackbody, a Comptonization model, the AMOL model, the cold gas model (i.e., HOT with $k_\mathrm{B}T=5\times10^{-4}$),  and the XABS  model.\\
The fit of 4U 1728-34 was produced using the following SPEX model components:
a disk blackbody, a Comptonization model, AMOL, the cold gas model, and XABS.\\
The fit of GX 340+00 was produced using the following SPEX models:
a blackbody, a power law, AMOL, the cold gas model, and XABS.
}
\end{table*}

\begin{figure}
 \begin{center}
 \includegraphics[scale=0.5]{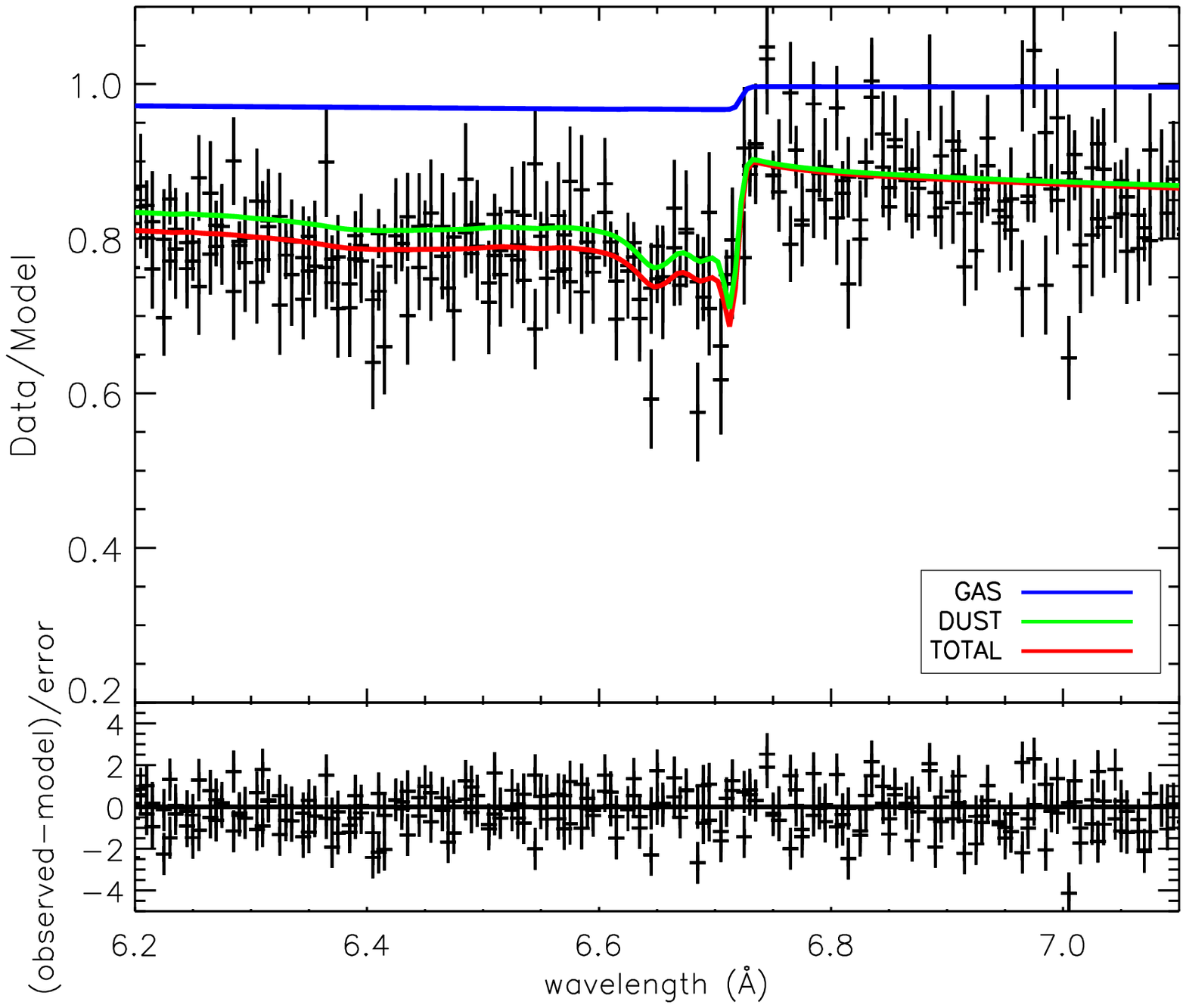}
 \caption{\small{Si K-edge GRS 1758-258.}}
 \label{fig:GRS_1758-258}
  \end{center} 
\end{figure}

\begin{figure}
 \begin{center}
 \includegraphics[scale=0.5]{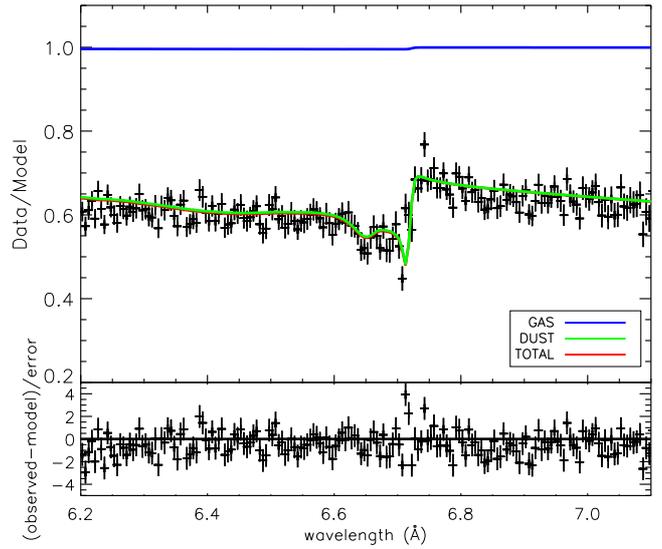}
 \caption{\small{Si K-edge GX 17+2.}}
 \label{fig:GX_17+2}
  \end{center} 
\end{figure}

\begin{figure}
 \begin{center}
 \includegraphics[scale=0.5]{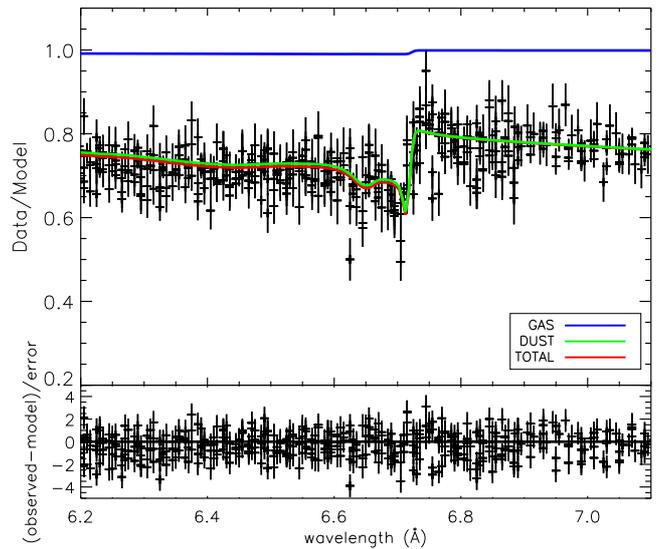}
 \caption{\small{Si K-edge 4U 1705-44.}}
 \label{fig:4U_1705-44}
  \end{center} 
\end{figure}

\begin{table*}
\caption{Best fit parameters for GRS 1758-258, GX 17+2, and 4U 1705-44}  % title of Table
\label{table:GRS1758258_GX172_4U170544} % is used to refer this table in the text
\centering % used for centering table
\begin{tabular}{l c c c c c c c} % centered columns (4 columns)
\hline\hline % inserts double horizontal lines
\rule{0pt}{2.3ex}
Source & \multicolumn{2}{c}{GRS 1758-258} & GX 17+2 &  \multicolumn{4}{c}{4U 1705-44} \\
Obsid  & 2429 & 2750 & 11088 & 5500 & 18086 & 19451 & 20082 \\ \hline  \noalign{\smallskip}% table heading
$N_\mathrm{H}^{\mathrm{cold}}$ ($10^{22}\,\mathrm{cm}^{-2}$)  & \multicolumn{2}{c}{$2.5\pm0.1$}& $2.0\pm0.1$ &\multicolumn{4}{c}{$2.0\pm0.1$}\\
$N_\mathrm{H}^{\mathrm{hot}}$ ($10^{20}\,\mathrm{cm}^{-2}$) & \multicolumn{2}{c}{$2.4^{+3.4}_{-1.2}$} & $1.4^{+1.2}_{-0.2}$ &  \multicolumn{4}{c}{$0.29^{+0.18}_{-0.23}$} \\
$k_\mathrm{B}T_{\mathrm{hot}}$ (keV) & \multicolumn{2}{c}{$1.3^{+0.5}_{-0.2}$} & $0.12\pm0.02$ & \multicolumn{4}{c}{$0.28^{+0.87}_{-0.18}$} \\
$k_\mathrm{B}T_{\mathrm{bb}}$ (keV) &$0.38\pm0.01$&$0.42\pm0.01$& - & - & - & - & - \\ 
$\Gamma_{\mathrm{pow}}$ & $3.9^{+0.1}_{-0.3}$ & $2.7\pm0.1$ & - & - &  - & - & -\\
$\mathrm{T}_{\mathrm{dbb}}$ (keV) & - & - & $1.7^{+0.2}_{-0.1}$  &  $0.42^{+0.07}_{-0.03}$ & $0.99^{+0.16}_{-0.33}$ & $0.53^{+0.18}_{-0.11}$ & $0.43^{+0.08}_{-0.07}$ \\ 
$k_\mathrm{B}T_{0\,\mathrm{comt}}$ (keV)& - & - & $0.60\pm0.01$ & $0.41\pm0.01$ & $0.65^{+0.05}_{-0.09}$ & $0.54\pm0.02$ & $0.49\pm0.01$\\ 
$k_\mathrm{B}T_{1\,\mathrm{comt}}$ (keV)& -  & - & $30^{+6}_{-11}$ &$21^{+20}_{-13}$&$47^{+39}_{-12}$&$40^{+53}_{-16}$&$18^{+22}_{-9}$\\   
$\tau_{\mathrm{comt}}$ &  - & - &$1.7^{+1.1}_{-0.6}$&$1.4^{+1.6}_{-1.3}$ & $0.9^{+1.8}_{-0.6}$ & $1.0^{+1.6}_{-0.7}$ & $1.6^{+1.2}_{-0.9}$\\
$F_{0.5-2\,\mathrm{keV}}\,(10^{-10}\mathrm{erg}\, \mathrm{cm}^{-2} \mathrm{s}^{-1})$ &$0.8\pm0.2$&$1.2\pm0.2$ &$4.3\pm0.8$&$1.4\pm0.1$&$3.6\pm0.4$&$2.3\pm0.2$&$2.3\pm0.2$\\
$F_{2-10\,\mathrm{keV}}\,(10^{-9}\mathrm{erg}\, \mathrm{cm}^{-2} \mathrm{s}^{-1})$ &$0.12\pm0.04$&$0.44\pm0.08$&$13\pm2$&$1.7\pm0.2$&$7.1\pm0.7$&$3.9\pm0.4$&$3.0\pm0.3$\\  \cline{2-3} \cline{5-8}
$C^2/\nu$& \multicolumn{2}{c}{4182/3770} & 1393/1109 & \multicolumn{4}{c}{5838/4459}  \\ 
\hline %inserts single line
\end{tabular}
\tablefoot{The fit of GRS 1758-258 was produced using the following SPEX model components:
a blackbody, a power law, AMOL, the cold gas model (i.e., HOT with $k_\mathrm{B}T=5\times10^{-4}$), and hot gas (HOT).\\
The fit of GX 17+2 was produced using the following SPEX model components:
a disk blackbody, a comptonization, AMOL, the cold gas model, and hot gas (HOT).\\
The fit of 4U 1705-44 was produced using the following SPEX model components:
a disk blackbody, a comptonization, AMOL, the cold gas model, and hot gas (HOT).
}
\end{table*}

\section{Si K-edge models}\label{si_models_ch3}

\begin{figure}
 \begin{center}
 \includegraphics[scale=0.5]{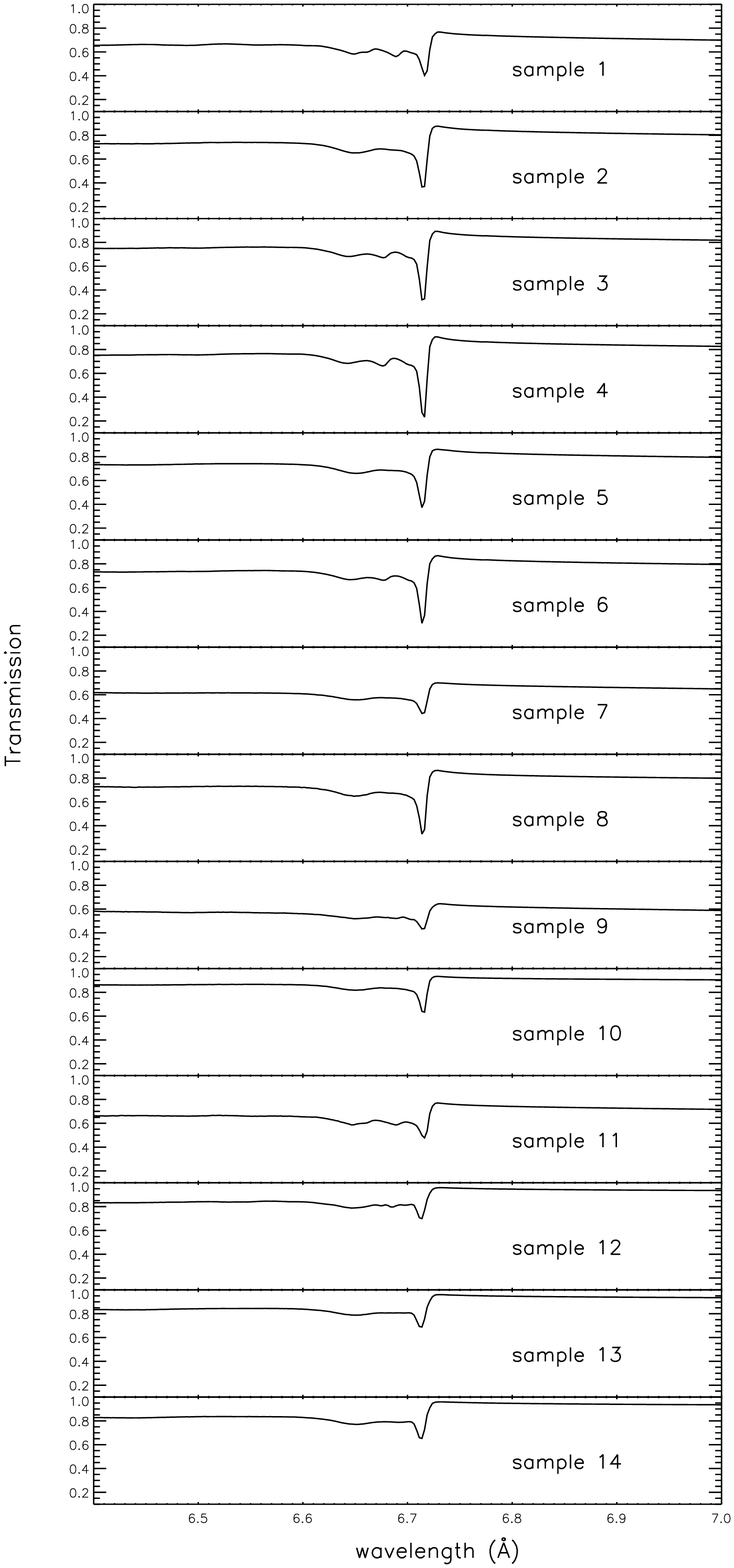}
 \caption{\small{Transmission of the six dust extinction models (absorption
and scattering) of the Si K-edge. 
The silicon column density has been set here to $10^{18}\,\mathrm{cm}^{-2}$
for all the dust models. Each model is indicated by a number  corresponding to the numbers in
Table~\ref{table:samples}.}}
 \label{fig:models}
  \end{center} 
\end{figure}

\begin{figure*}
 \begin{center}
 \includegraphics[scale=0.9]{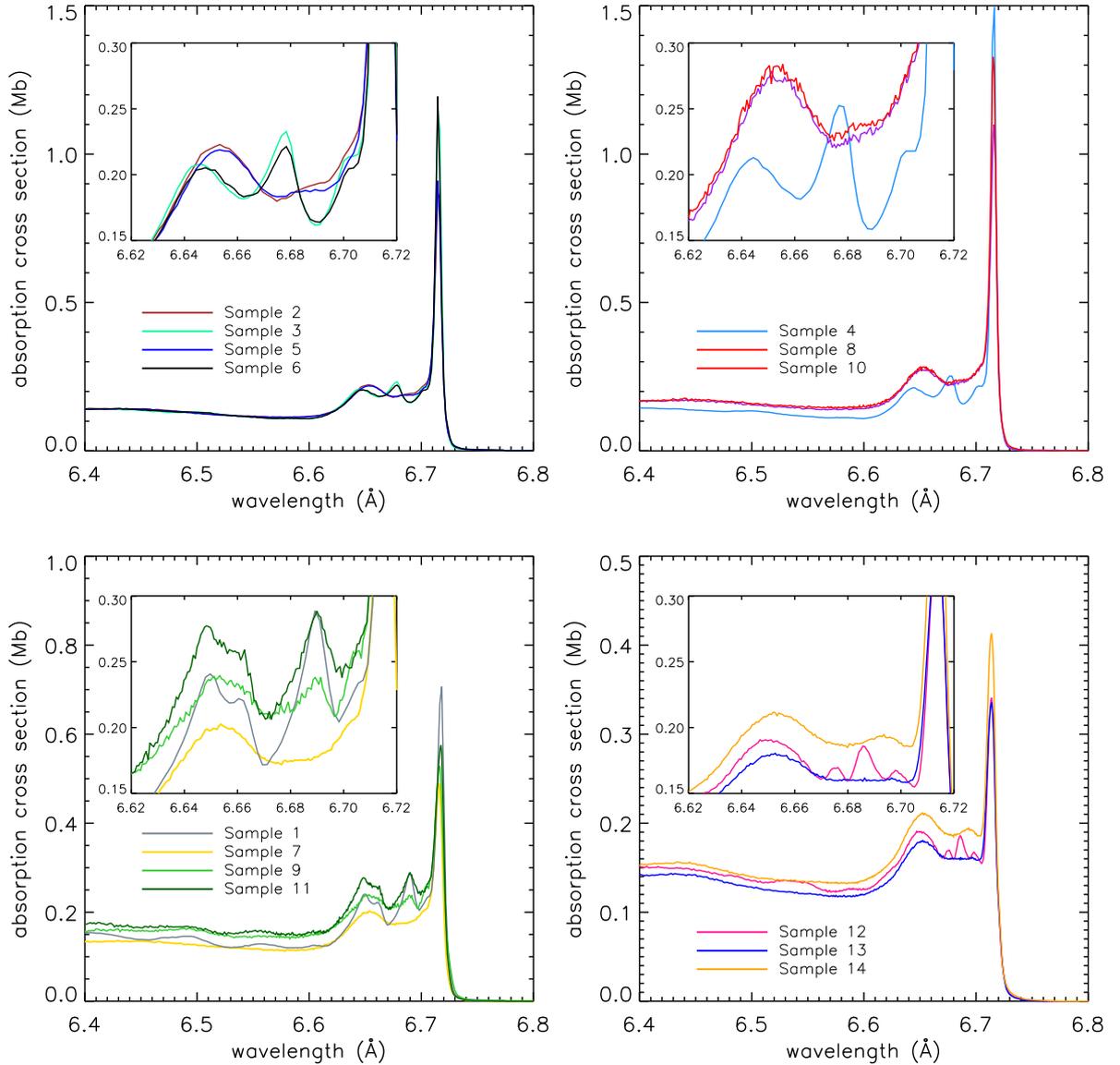}
 \caption{\small{The Si K-edge of dust samples 1-14 (see Table~\ref{table:samples}). The x-axis shows the energy in $\AA$ and the 
 y-axis shows the amount of absorption indicated by the cross section (in Mb per Si atom). 
 The samples are shown in different panels for easier comparison. 
 The top left panel shows the difference between amorphous and crystalline pyroxenes (samples 2, 3, 5, and 6). 
 The top right panel shows the difference between amorphous and crystalline enstatite (samples 4 and 10), including  
 an amorphous pyroxene. The bottom left panel focusses on the olivine dust samples and the bottom right panel shows our three quartz samples.}}
 \label{fig:edges_again}
  \end{center} 
\end{figure*}

Figure~\ref{fig:models} shows the extinction profiles around the Si-K edge for the compounds 1 to 14 (see Table~\ref{table:samples}).
All the profiles are implemented in the AMOL model of the spectral 
fitting code SPEX. The absolute cross sections of the models
used in this analysis are available in tabular form\footnote{\tiny{\url{www.sron.nl/~elisa/VIDI/}}}.
Furthermore, we show the laboratory edges from Figure~\ref{fig:all_edges} 
in individual panels for comparison 
in Figure~\ref{fig:edges_again}.  

 \end{appendix}

%-------------------------------------------------------------------

\bibliography{zeegers2}

\end{document}